%% file: schiff_complex.tex
\documentclass[twoside,twocolumn,9pt]{article}
\usepackage{extsizes}
\usepackage[super,sort&compress,comma]{natbib} 
\usepackage[version=3]{mhchem}
\usepackage[left=1.5cm, right=1.5cm, top=1.785cm, bottom=2.0cm]{geometry}
\usepackage{balance}
\usepackage{mathptmx}
\usepackage{sectsty}
\usepackage{graphicx} 
\usepackage{lastpage}
\usepackage[format=plain,justification=justified,singlelinecheck=false,font={stretch=1.125,small,sf},labelfont=bf,labelsep=space]{caption}
\usepackage{float}
\usepackage{fancyhdr}
\usepackage{fnpos}
\usepackage[english]{babel}
\addto{\captionsenglish}{%
  
}
\usepackage{array}
\usepackage{droidsans}
\usepackage{charter}
\usepackage[T1]{fontenc}
\usepackage[usenames,dvipsnames]{xcolor}
\usepackage{setspace}
\usepackage[compact]{titlesec}
\usepackage{hyperref}
\usepackage{subcaption} 
\usepackage{placeins} 

\title{Vibrational Circular Dichroism enhancement in conformationally flexible transition metal complexes}

\author{
\begin{minipage}{\textwidth}
	Mariia Sapova,\textsuperscript{a} Menno de Boer,\textsuperscript{a, b} Wybren Jan Buma,*\textsuperscript{c, d} Lucas Visscher,*\textsuperscript{a} 
\end{minipage}
}

\newcommand{\affiliation}{
\begin{itemize}	    

\item[a] M. Sapova,  M. de Boer, Prof. Dr. L. Visscher*\\
Department of Chemistry and Pharmaceutical Sciences, Faculty of Sciences, Vrije Universiteit Amsterdam, De Boelelaan 1108, 1081 HZ Amsterdam, The Netherlands \\
E-mail: L.Visscher@vu.nl

\item[b] M. de Boer\\
Theoretische Organische Chemie, Organisch-Chemisches Institut, Center for Multiscale Theory and Computation (CMTC), University of Münster, 48149 Münster, Germany \\

\item[{c}]Prof. Dr. W.J. Buma*\\
Van’t Hoff Institute for Molecular Sciences, University of Amsterdam, Science Park 904,
1098 XH Amsterdam, The Netherlands \\
E-mail: W.J.Buma@uva.nl\\

\item[{d}] Prof. Dr. W.J. Buma*\\
Radboud University, Institute for Molecules and Materials, FELIX Laboratory, Toernooiveld 7c, 6525 ED Nijmegen, The Netherlands \\

\end{itemize}}

\begin{document}

\maketitle
\affiliation

\begin{abstract}
We extend our previously developed approach for calculating enhanced vibrational circular dichroism (VCD) spectra of transition-metal complexes to systems for which conformational flexibility is key to reproduce and elucidate experimental spectra. Treating both the Gibbs free energies and electronic excitation energies as fitting parameters, we show for Co(II){bis[N-(1-arylethyl)- salicylaldiminato]} Schiff base complexes that these calculations can excellently reproduce the experiment. An important conclusion is that the DFT-optimized conformer set is spectrally redundant and that the model can be reduced from 18 conformers to only two conformers with opposite chirality at the metal center ($\Lambda$/$\Delta$) without affecting the agreement between theory and experiment, both with respect to the VCD spectrum as well population distribution over the $\Lambda$ and $\Delta$ conformers. Finally, we numerically confirm the symmetry-selective nature of the enhancement and formulate the corresponding selection rules within the framework of the pertaining Sum-Over-States expressions.
\end{abstract}

\section{Introduction}
Vibrational circular dichroism (VCD)\cite{Nafie_2008_review, Stephens_2008_review, KUROUSKI2017} spectroscopy, the differential absorption of left- and right-handed circularly-polarised infrared light, has become one of the most powerful spectroscopic methods to determine the absolute configuration and conformational distribution of chiral molecules in solution. However, its application and interpretation are often complicated by the inherent small signal intensities. Enhancements of VCD signals as observed experimentally for open-shell transition metal complexes~\cite{Pescitelli2024_review} have therefore attracted major interest as a fundamental understanding of the underlying mechanisms might very well pave the way for a further tailoring of applications of VCD spectroscopy. 

In these open-shell transition metal systems, VCD signals are significantly amplified relative to a closed-shell isostructural reference, typically the corresponding Zn(II) complex. This enhancement is commonly attributed to the presence of low‐lying electronic excited states (LLES) that couple electronic and vibrational transitions and can increase the VCD intensity by up to two orders of magnitude ~\cite{Nafie_2004, Nafie2011VOA}. A recent report on Co(II) complexes with two diketonate and one bis(oxazoline) ligand~\cite{Papi2025} showed especially large VCD dissymmetry factors, up to  $5 \times 10^{-2}$. Even though most studies focus on open-shell Co systems, enhanced VCD has also been reported for Ni, Cu, Ru, Cr, Rh, Ir, Fe, and V complexes with a variety of chiral ligand environments, including diamines, amino-acid and peptide derivatives, $\beta$-diketonates, Schiff bases, helicene-based ligands, and polypyridyl frameworks~\cite{Pescitelli2024_review}. From these studies it has become clear that enhanced VCD can serve as a sensitive probe of local chiral structure ~\cite{Domingos2014SwitchableAmplification, Domingos2014AmplifiedBiomolecular, Berardozzi2016CoInducedGiantVCD}. 

Putting such enhancements to practical use has proven to be quite difficult if not impossible as a reliable computational approach to analyze and predict the observed enhanced VCD spectra has for a long time been lacking. Recently, we introduced a novel methodology based on the sum-over-states approach introduced by Nafie~\cite{Nafie_2004, Nafie2011VOA} to calculate enhancements from low‑lying $d$–$d$ transitions in the Co(II) and Ni(II) (–)-sparteine‑Cl$_2$ complexes ~\cite{Sapova2026}. For this relatively rigid sparteine system~\cite{WIBERG2000239}, we showed that the limited accuracy of the calculated excitation energies hinders a direct prediction of the VCD intensity corrections. However, by introducing a practical strategy to mitigate this uncertainty~\cite{Sapova2026} we demonstrated that it was possible to quantitatively reproduce enhanced experimental spectra and obtain simulated VCD similarity scores above 0.4, a threshold considered reliable for an absolute configuration assignment.
 
In 2018 Pescitelli  \textit{et al.} ~\cite{Pescitelli2018_schiff} reported a chiroptical superspectrum of Co(II){bis[N-(1-arylethyl)- salicylaldiminato]} Schiff base complexes. Unlike most molecular systems, these Co(II) complexes display an almost continuous set of CD bands spanning the UV to IR regions. In the 1800–3000 cm$^{-1}$ region, broad CD bands were observed that were assigned to low‑lying $d$–$d$ transitions while in the fingerprint region an approximate tenfold enhancement of bands was observed for the Co(II) complex relative to the isostructural Zn(II) and Cu(II) analogues~\cite{Zn_schiff, Cu_schiff}. The remarkable feature of the VCD spectra of these complexes is that most of the fingerprint bands are negative, leading to a spectrum that is almost monosignate. This contrasts with the usual alternating sign patterns observed for homoleptic Zn(II) and Cu(II) complexes. A similar phenomenon was reported for square-planar -(R, R)-N,N'-Bis(3,5-di-tert-butylsalicylidene)-1,2-cyclohexanediamine with Co(II)~\cite{Salen_Alshalalfeh}.
This unusual monosignate appearance was later interpreted as a symmetry-dependent enhancement effect~\cite{Pescitelli2019}. It was suggested that for complexes with $C_2$ symmetry  normal modes of B symmetry are selectively enhanced and inverted in sign through coupling with low-lying electronic states of matching symmetry. Tome\v{c}ek and Bou\v{r} subsequently included electronic-vibrational coupling beyond the standard magnetic-field perturbation treatment and reproduced some enhancement effects, but the monosignate VCD pattern was not fully captured ~\cite{Tomecek2020}.


In the present work, we revisit the Co(II) salicylaldiminato system using our enhanced VCD framework and test whether the unusual Co(II) fingerprint region in the VCD spectrum can be described quantitatively. Compared to our previous work on sparteine complexes, the Co(II) salicylaldiminato complex provides a more demanding test case because it is conformationally flexible.  For flexible molecules, the spectrum should be averaged over multiple conformers using Boltzmann weights derived from Gibbs free energies. However, the relative conformer energies depend on the chosen level of theory and  their uncertainties can strongly affect conformer populations and averaged VCD spectra~\cite{Koenis2019Taming, Nicu_artificial}. Here we account for the uncertainty in both conformer Gibbs energies as well as calculated excitation energies. We show that the standard Magnetic Field Perturbation (MFP) approach used to calculate VCD spectra~\cite{Stephens1985VCDTheory, Stephens1996AATDFT} reproduces neither the intensity nor the shape of the experimental spectrum, whereas inclusion of enhancement by the lowest electronically excited state and using only two conformers of opposite metal center chirality leads to very good agreement between theory and experiment. We also find that the optimized conformer populations indicate a substantially larger contribution from the $\Delta$ conformer than predicted from the DFT-calculated Boltzmann weights, which is in fact in better agreement with experimental observations. Finally, an analysis using approximate $C_2$ symmetry shows that the enhancement is mainly due to B-like vibrational modes, providing a numerical confirmation of the proposed symmetry-dependent mechanism for realistic, non-ideal conformer geometries.

The paper is organized as follows. In the \nameref{sec:theory} section we briefly describe the enhanced VCD formalism and discuss the symmetry-dependent selection rules expected for an idealized $C_2$-symmetric complex. Subsequently we describe in the ~\nameref{sec:comp_details} section the conformational search and DFT reoptimization. In the \nameref{sec:res_and_dis} section we then select the relevant conformers and compare MFP-VCD and enhanced VCD results with experiment. Finally, we introduce an approximate $C_2$ symmetry analysis of the relevant conformers and discuss the implications of this analysis for the nearly monosignate fingerprint region of the VCD spectrum.

\section{Theoretical background}\label{sec:theory}

\subsection{Enhanced VCD}

To calculate the enhancement of VCD signals in a flexible complex, we determine the set of $N$ lowest-energy conformers and introduce  uncertainties in their relative ground-state energies and their electronic excitation energies. For each conformer enhanced VCD spectra are computed using a perturbative correction to MFP-calculated rotational strengths as described by  Nafie's vibronic coupling formalism~\cite{Nafie_2004, Nafie2011VOA}. The detailed derivation is given in Sapova \textit{et al.}~\cite{Sapova2026} equations 1-16.

\begin{equation}
R_i^{\textup{full}}= \textup{Im} [\underbrace{\mathbf{E_i}^{\textup{tot}} \cdot {\mathbf{M_i}}^{\textup{tot}}}_{\textup{MFP}} + \underbrace{\mathbf{E}_i^{\textup{tot}} \cdot \mathbf{M}_i^{\textup{enh}} + \mathbf{E}_i^{\textup{enh}} \cdot \mathbf{M}_i^{\textup{tot}}}_{\textup{enhancement correction}}]
\label{eq:rot_strength}
\end{equation}

The electric and magnetic dipole transition moments (EDMT and MDTM) in the MFP formalism are defined as:
\begin{equation} \label{eq:EDTM_MDTM}
\begin{aligned}
E^{\textup{tot}}_{\beta,i} &={\Bigg({\frac{\hbar}{{\omega}_{i}}}\Bigg)}^{\frac{1}{2}} \sum_{\lambda\alpha}\Bigg[{\frac{\partial \langle \Psi_{g}(\mathbf{R}) | \mu_{\beta}| \Psi_{g}(\mathbf{R})\rangle }{\partial R_{\lambda\alpha}}\Bigg | }_{\mathbf{R}=\mathbf{R}_{0}} + \\ &+ e Z_{\lambda} {\delta}_{\alpha\beta}\Bigg]S_{\lambda\alpha,i} \\
M^{\textup{tot}}_{\beta,i} & = -{({2{\hbar}^{3}{\omega}_{i}})}^{\frac{1}{2}} \sum_{\lambda\alpha}\Bigg[{\Bigg \langle \frac{\partial \Psi (\mathbf{R},\mathbf{B})}{ \partial R_{\lambda \alpha}} \Bigg| \frac{\partial \Psi(\mathbf{R},\mathbf{B})}{ \partial B_{\beta}}    \Bigg \rangle \Bigg |}_{\mathbf{R}_{0}, \mathbf{B}=0}  \\ &+ i \frac{e Z_{\lambda}}{4 \hbar c} \sum_{\gamma} {\epsilon}_{\alpha \beta \gamma} R_{\lambda \gamma}^{0} \Bigg]S_{\lambda\alpha,i}
\end{aligned}
\end{equation}

\noindent where $\mathbf{S}$ is the transformation matrix from Cartesian ($\lambda\alpha$) to normal ($i$) nuclear coordinates, $\omega_{i}$ the harmonic angular frequency of the $i^{th}$ mode, $\lambda$ indexes nuclei and $\alpha$ and $\beta$ denote Cartesian components. The two terms in the big square brackets of the EDMT and MDTM expressions are associated with, respectively, the electronic and the nuclear contribution to the Atomic Polar Tensor (APT) and the Atomic Axial Tensor (AAT). The enhancement corrections to EDTM and MDTM are then given by
\begin{equation}
\label{enhancement}
\begin{aligned}
& E_{i,\beta}^{\textup{enh}} = 2 {\Bigg({\frac{\hbar}{{\omega}_{i}}}\Bigg)}^{\frac{1}{2}} \sum_{e\neq g} \frac{\omega_{i}^{2}}{\omega_{eg}^{2} - \omega_{i}^{2}} \Bigg [  \langle {\Psi_g} |  \mu_{\beta}|  \Psi_e \rangle \langle \Psi_{e} |  \frac{\partial \Psi_{g}}{\partial Q_{i}} \rangle\Bigg ] \\
& M_{i,\beta}^{\textup{enh}} = - {({2{\hbar}^{3}{\omega}_{i}})}^{\frac{1}{2}} \sum_{e\neq g} \frac{\omega_{i}^{2}}{\omega_{eg}^{2} - \omega_{i}^{2}} \Bigg [  
\frac{\langle {\Psi_g} |  m_{\beta}|  \Psi_e \rangle }{\omega_{eg}}
\langle \Psi_{e} |  {\frac{\partial \Psi_{g}}{\partial Q_{i}} \rangle\Bigg ]} 
\end{aligned}
\end{equation}

\noindent where $\omega_{eg}$ are electronic excitation energies and $\langle \Psi_{e} |  {\frac{\partial \Psi_{g}}{\partial Q_{i}}} \rangle$ are the non-adiabatic couplings transformed to a normal mode basis.

\FloatBarrier

\subsection{Symmetry-dependent enhancement}

We want to examine which terms in ~\autoref{eq:rot_strength} are allowed by symmetry. To this purpose we consider a complex with $C_2$ symmetry, but note that this analysis can also easily be extended to other point groups. The $C_2$ character table is shown in \autoref{tab:C2_character} with the $z$ axis chosen along the $C_2$ axis; $z$ therefore transform as A, while $x$ and $y$ transform as B.

\begin{table}[]
\centering
\caption{Character table of $C_2$ point group}
\begin{tabular}{cccc}
  & E & $C_2$ & \begin{tabular}{c} linear functions\\ rotations\end{tabular} \\
\hline
A & 1 & 1  & $z$, $R_z$ \\
B & 1 & -1 & $x$, $y$, $R_x$, $R_y$                                   
\end{tabular}
\label{tab:C2_character}
\end{table}

We first examine which Cartesian components contribute to the MFP rotational strength $R_i$ for vibrational modes of different symmetry. To this purpose we express the APT and AAT in the normal mode basis. For the APT the symmetry-allowed components are determined by the direct product of the irreducible representations of the electric dipole component and the normal mode:
\begin{equation}
\begin{aligned}
\frac{\partial \langle \Psi_{g}| \mu_{\beta}| \Psi_{g}\rangle }{\partial Q_{i}} 
= \sum_{\lambda \alpha}\frac{\partial \langle \Psi_{g}(\mathbf{R}) | \mu_{\beta}| \Psi_{g}(\mathbf{R})\rangle}{\partial{R_{\lambda\alpha}}}S_{\lambda\alpha,i}
\\ \Gamma(\frac{\partial \langle \Psi_{g}| \mu^E_{\beta}| \Psi_{g}\rangle }{\partial Q_{i}}) = \Gamma(\mu^E_{\beta}) \otimes \Gamma(Q_{i})
\end{aligned}
\end{equation}
Thus, A modes couple to the $\mu_z$ component, whereas B modes couple to the $\mu_x$ and $\mu_y$ components. The same argument applies to the AAT. For point groups that contain no improper symmetry operations, \textit{i.e.}, for all chiral point groups, rotations transform in the same way as linear functions.
\begin{equation}
\Gamma\langle\frac{\partial \Psi_{g}}{\partial Q_i}|\frac{\partial \Psi_{g}}{\partial B_\beta}\rangle = \Gamma(Q_i) \otimes \Gamma(B_\beta)
\end{equation}

As a result, for $C_2$ the MFP rotational strength
$ R_i = \sum_{\beta} E_{\beta i} M_{\beta i}$
has different nonzero Cartesian contributions depending on the symmetry of the vibrational mode.
For A modes, only the $z$ component is symmetry allowed, whereas for B modes both the $x$ and $y$ components are allowed.
\begin{equation}
\begin{aligned}
R_i^{A} & = E_{zi}M_{zi} \\
R_i^{B} & = E_{xi}M_{xi} + E_{yi}M_{yi}    
\end{aligned}
\end{equation}

We now apply the same symmetry analysis to the enhancement terms. 
Enhanced EDTM and MDTM  depend on two quantities: the electronic dipole transition moments (electric and magnetic) and non-adiabatic couplings.
The symmetry of the transition dipole moments ($x = {\mu, m}$) is given by
\begin{equation}
\Gamma(\langle {\Psi_g} |x_{\beta}| \Psi_e \rangle) = \Gamma(\Psi_g) \otimes \Gamma (x_{\beta}) \otimes \Gamma(\Psi_e) 
\end{equation}
The non-adiabatic coupling term can also be written in the normal mode basis and is given by
\begin{equation}
\langle \Psi_{e} |  \frac{\partial \Psi_{g}}{\partial Q_{i}} \rangle = \sum_{\lambda\alpha} \langle \Psi_{e} |  \frac{\partial \Psi_{g}}{\partial R_{\lambda\alpha}}\rangle S_{\lambda\alpha,i}
\end{equation}
This coupling is only nonzero when the normal mode transforms as the direct product of the ground- and excited-state irreducible representations:
\begin{equation}
\Gamma(Q_i) = \Gamma(\Psi_g) \otimes \Gamma (\Psi_e)
\end{equation}

The resulting selection rules are summarized in \autoref{tab:C2_rules}. It indicates which vibrational modes can be enhanced by a given electronic transition and through which Cartesian component of the magnetic dipole. Vibrational modes of different symmetry are enhanced by different electronic transitions. The lowest $d$-$d$ transition is nearly electric dipole forbidden for the salicylaldiminato complex. As a result, the enhancement correction to the EDMT in ~\autoref{enhancement} are negligible. For the enhancement correction to the MDTM in ~\autoref{enhancement} 
an electronic A-symmetry transition enhances A-vibrational modes through the component $m_z$, whereas an electronic B-symmetry transition enhances vibrational modes of B symmetry through the $m_x$ and $m_y$ components. Enhancement selection rules for other chiral point groups are collected in Table S3. 

The present analysis provides a more rigorous description of the symmetry-selective enhancement proposed by Pescitelli et al.~\cite{Pescitelli2019}. 
The symmetrical Co(II) salicyldiminato complex that is studied here
has an A-symmetry ground state and three lowest excited states of B symmetry.
Transitions between these states are thus B transitions and according to \autoref{tab:C2_rules} are therefore expected to enhance B modes through the dipole components of the magnetic transition $m_x$ and $m_y$.

\begin{table}[]
\caption{Symmetry-allowed magnetic dipole enhancement contributions in the $C_2$ point group.}
\begin{tabular}{cccc}
Transition     & Enhanced mode & $m$ component & $\mu$ component \\
\hline
A $\rightarrow$ A  & A             & $m_z$   & $\mu_z$          \\
A $\rightarrow$ B  & B             & $m_x$, $m_y$    & $\mu_x$, $\mu_y$   \\
B $\rightarrow$ A  & B             & $m_x$, $m_y$     & $\mu_x$, $\mu_y$  \\
B $\rightarrow$ B  & A             & $m_z$        & $\mu_z$ 
\end{tabular}

\label{tab:C2_rules}
\end{table}

\section{Computational details}\label{sec:comp_details}

All calculations were performed with ORCA 6.0~\cite{ORCA_6} assuming a high-spin state of the complex. We carried out the conformational search using the GOAT algorithm~\cite{GOAT} with GFNn-xTB~\cite{Grimme2017GFN1xTB} and then reoptimized all conformers within a 20 kcal/mol energy window at the B3LYP~\cite{becke1993density}-D3BJ~\cite{Grimme2011D3BJ}/def2-TZVP~\cite{Weigend2005Def2} level with a CPCM~\cite{Barone1998CPCM, Cossi2003CPCM} solvent model (CHCl$_3$ solvent). The starting structures were taken from X-ray data for the corresponding Zn(II) complex~\cite{Zn_schiff}. Throughout the main text and SI we use the GOAT conformer labeling based on their GFNn-xTB energies. 

IR, MFP-VCD and TDDFT calculations reported in the main text use the same level of theory. Enhanced VCD spectra were calculated using our in-house Python implementation of Nafie’s vibronic coupling theory~\cite{eVCD} and use the experimental data kindly provided by Prof. Pescitelli  (VCD and IR data published in ~\cite{Pescitelli2018_schiff}). To assess the agreement between theory and experiment, we used Shen's~\cite{Shen2010SimIRVCD} simIR and simVCD similarity metrics. For both IR and VCD calculations, we used a frequency scaling factor obtained by maximizing SimIR between calculated and experimental IR spectrum of the lowest-lying conformer.

\section{Results and discussion}\label{sec:res_and_dis}

\subsection{Co(II) complex structure}

The molecule of interest is a high-spin pseudotetrahedral Co(II) complex with two coordinated bidentate Schiff base ligands. Pesticelli \textit{et al.}~\cite{Pescitelli2018_schiff} reported absorption and CD superspectra for Co(II){bis[N-(1-arylethyl)-salicylaldiminato]} with different aryl groups. Here, we focus in the main text on the $p$‑ClC$_6$H$_4$ derivative. The spectra for other \textit{para}‑phenyl substituents differ mainly in the low intensity noisy 1200–1000 cm$^{-1}$ region and are provided in the SI (Figures S8-S10).

\begin{figure}[h]
    \centering
    \includegraphics[width=8.3cm]{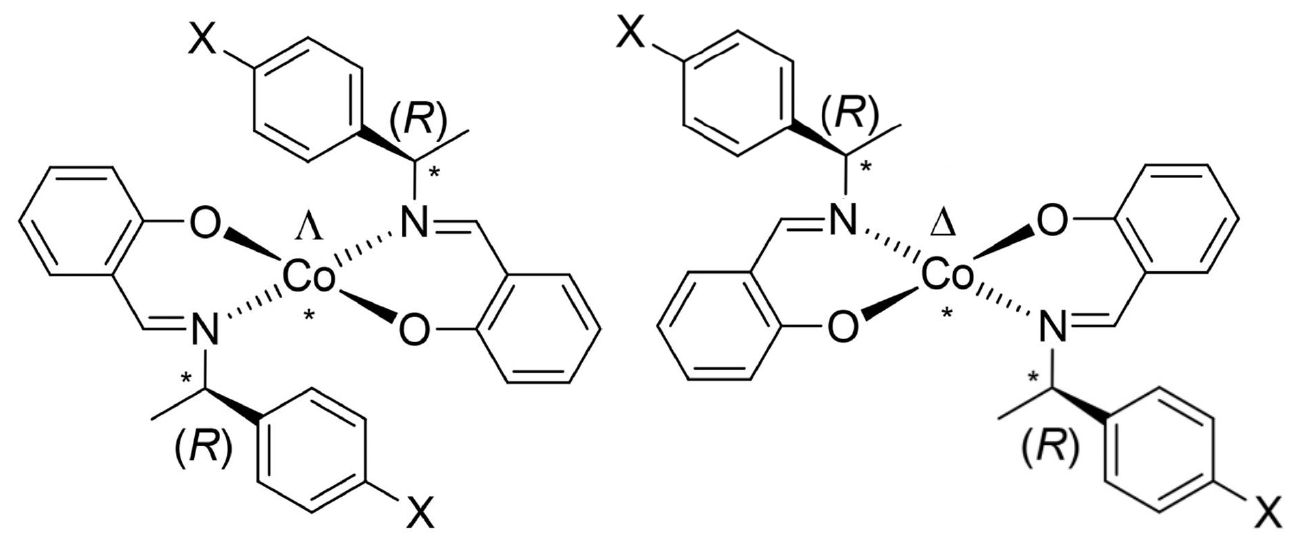}
    \caption{(R)-Co(II)bis[N-(1-arylethyl)- salicylaldiminato $\Lambda$ and $\Delta$ chirality.}
    \label{fig:scheme}
\end{figure}

The complex exhibits metal center chirality, which is denoted using the $\Delta/\Lambda$ notation. In the solid state ($R$) and ($S$) ligands form uniquely $\Lambda$ and $\Delta$ complexes, respectively, whereas in solution an equilibrium is established. Here, we calculate the enhanced VCD spectrum for the ($R$) isomer (~\autoref{fig:scheme}). Variable temperature ECD measurements were used to identify the equilibrium between the $\Lambda$ and $\Delta$ conformers. Although Pesticelli \textit{et al.} noted that the number of data points was limited for an accurate determination, Boltzmann fitting of CD intensities at 380 nm yields a Gibbs free‑energy difference of ~0.8 kcal/mol between the $\Lambda$ and $\Delta$ diastereomers~\cite{Pescitelli2018_schiff}. This corresponds to a ratio of approximately 80:20 of $\Lambda$‑($R$)‑Co to $\Delta$‑($R$)‑Co at 300 K (see ~\autoref{fig:scheme}), which is similar to NMR results for the Zn(II) analogue.

\subsection{Enhanced VCD calculation}

The central question in the enhanced VCD calculation was how to choose an appropriate set of conformers. The GOAT~\cite{GOAT} search for the $\Lambda$ structure produced 26 conformers within an 8.6 kcal/mol energy window, whereas the corresponding $\Delta$ search yielded as many as 51 conformers.
The subsequent DFT reoptimization resulted in 8 $\Lambda$-(R) and 10 $\Delta$-(R) geometrically unique conformers within a 6 kcal/mol energy window (Table S1). Within the same metal chirality, these optimized structures differ mainly by rotations of the aryl groups around the bridging carbon (Figures S1 and S2). To identify unique conformers, we computed RMSDs over symmetry-equivalent atom mappings using the RDKit Python library followed by clustering using a heavy‑atom RMSD threshold of 0.125 Å~\cite{Pracht}. 

In contrast to previously reported B3LYP calculations~\cite{Pescitelli2018_schiff, Pescitelli2019}, which predicted one of the $\Delta$ conformations to have the lowest energy, our calculations find $\Lambda$-(R)-conf 11 as the lowest‑energy structure. This is consistent with the experimentally observed preference for $\Lambda$ metal center chirality. Moreover, we do not find a strong preference towards one particular $\Lambda$ conformer as was reported in the previous calculations, but find several $\Lambda$ conformers that are close in energy, with five conformers having Boltzmann weights larger than 0.05. We attribute these differences to the fact that the present geometry optimization includes both dispersion correction and a continuum solvent model. 

The lowest energy $\Delta$ conformer ($\Delta$-(R)-conf 0) lies 2.5 kcal mol$^{-1}$ above $\Lambda$-(R)-conf 11, giving a DFT $\Lambda$:$\Delta$ ratio of approximately 0.99:0.01. This differs from experimental estimates, where the $\Delta$ contribution is much larger. We therefore include uncertainty in the computed Gibbs free energies to account for DFT errors and obtain a more accurate conformer population model.

\begin{figure}
    \centering
    \includegraphics[width=9cm]{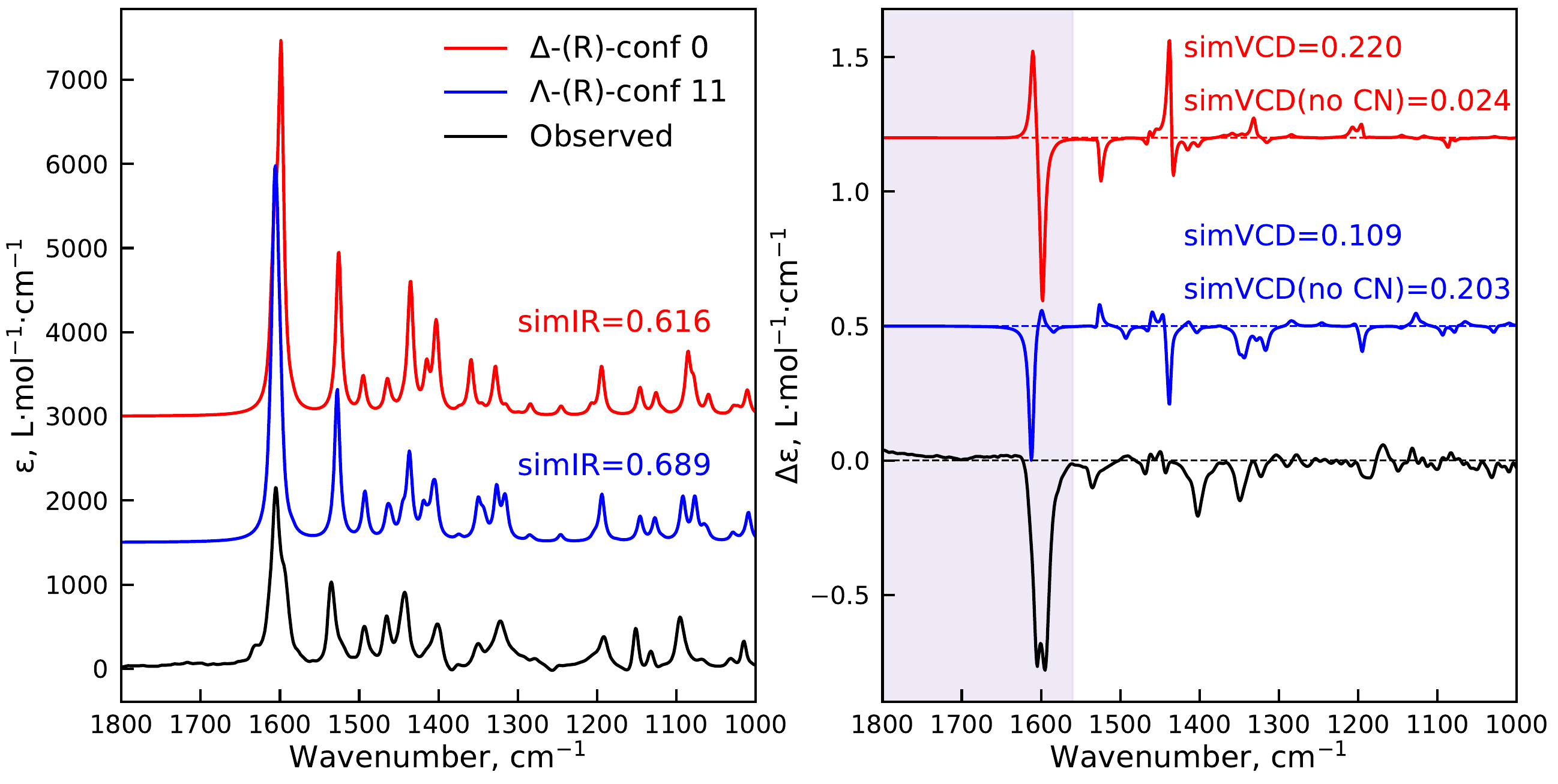}
    \caption{IR (left) and MFP-VCD (right) calculated spectra for lowest energy $\Lambda$ and $\Delta$ conformers. SimVCD values are calculated over the full 1800-1000 cm$^{-1}$ range, simVCD (No CN) excludes the shaded C=N stretching region.}
    \label{fig:ir_mfp}
\end{figure}

\begin{figure*}[h]
\centering
\includegraphics[width=18cm]{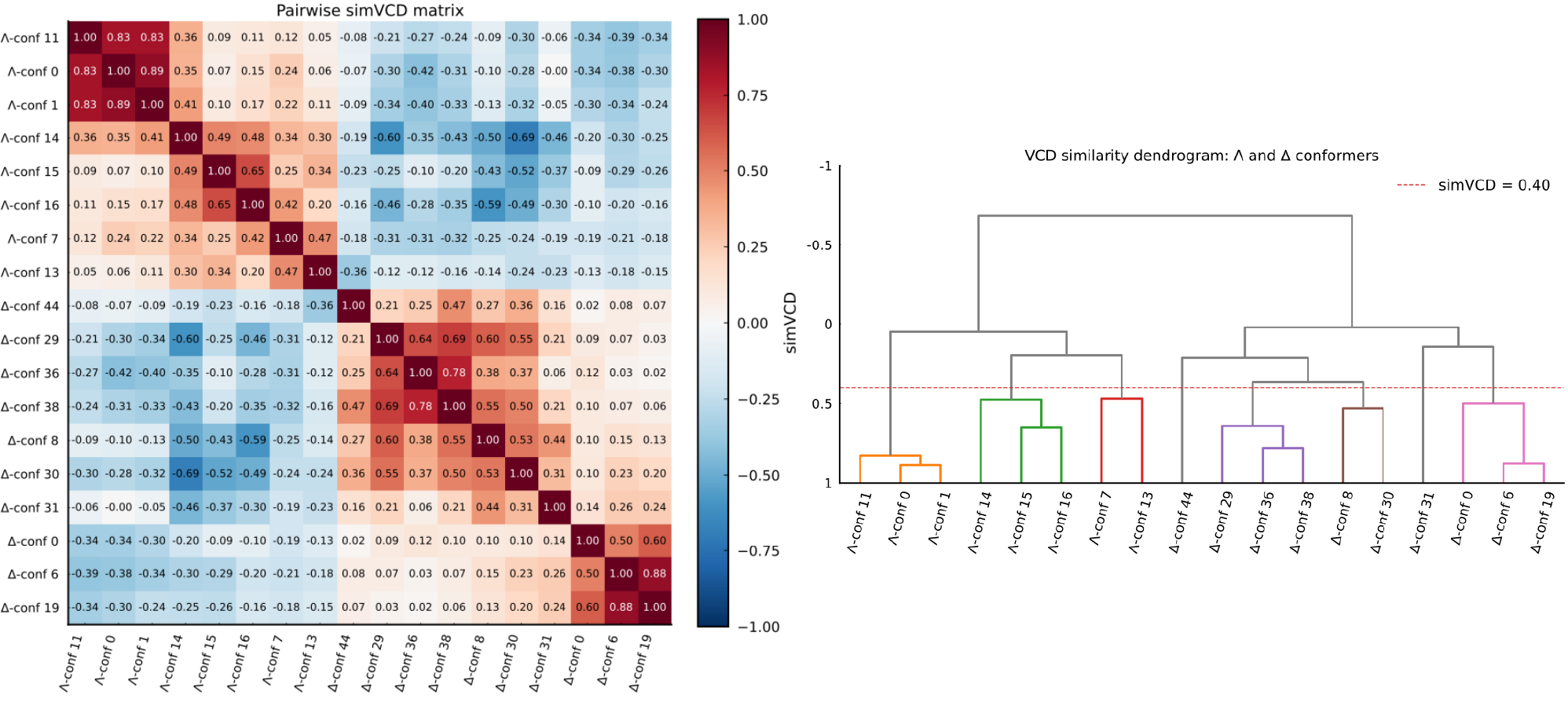}
\caption{SimVCD pairwise matrix for spatially unique conformers in dendrogram order (left) and hierarchical clustering of simulated conformer spectra (right).}
\label{fig:dendrogram_heatmap}
\end{figure*}

We first calculated IR and MFP-VCD spectra for all unique conformers. In the main text we show spectra for the lowest-energy $\Lambda$ and $\Delta$ conformers (~\autoref{fig:ir_mfp}), spectra for the other conformers are given in Figures S3 and S4. The calculated IR spectra are only weakly affected by conformational details and reproduce the experimental spectrum quite well with simIR>0.6. In contrast, the MFP-VCD spectra depend strongly on the chirality of the metal center with the $\Lambda$ and $\Delta$ conformers giving approximately opposite features, particularly in the 1700-1400 cm$^{-1}$ region.  Unlike previously reported results~\cite{Pescitelli2019}, the lowest-energy $\Lambda$ conformer already shows mostly negative features in MFP-VCD spectra, especially around and below 1500 cm$^{-1}$. The overall similarity metric remains, however, low as it is dominated by intense peaks in 1700-1500 cm$^{-1}$ region. However, even though excluding the 1600 cm$^{-1}$ region improves simVCD, MFP-VCD does not lead to values that are considered as reliable.  

A more detailed inspection of the VCD spectra reveals that within a given metal center chirality some conformers -for example $\Lambda$-conf 0, $\Lambda$-conf 1 and $\Lambda$-conf 11- produce very similar VCD spectra, whereas others such as $\Lambda$-conf 1 and $\Lambda$-conf 7 show more pronounced differences (Figure S4). Conformers with highly similar spectra cannot be assigned with independent weights from the experimental VCD spectrum and will therefore introduce redundant parameters into the enhanced VCD fitting. To identify spectrally redundant conformers, we computed the pairwise simVCD matrix for the 18 conformers and clustered them using a threshold of simVCD>0.4 (~\autoref{fig:dendrogram_heatmap}). This procedure divided the conformer pool into 8 spectral clusters of which we retained for each the lowest-energy conformer. 

To include the effects of coupling to electronically excited states, we calculated the enhancement corrections to the electric and magnetic dipole transition moments using TD-B3LYP-D3BJ/def2-TZVP/CPCM transition dipole moments and non-adiabatic couplings. In our previous work~\cite{Sapova2026} this level of theory gave good agreement with the CASSCF reference. We recall that one of the conclusions of this work was that enhancements are extremely sensitive to the excitation energies, making it challenging for quantum chemistry methods to predict these values with sufficient accuracy. We therefore treated these key excitation energies as adjustable parameters and found that in this way we could obtain excellent agreement between experimental and computed spectra. In the present case there is a substantial gap between the first electronically excited state and higher ones, which allows us to include only one state in the excitation energy optimization procedure. The first excitation energy varies only weakly for the conformers: the mean energy over all 18 conformers is 0.796 eV with a standard deviation of 0.022 eV (see Table S2). We therefore used a single shared excitation energy parameter rather than independent excitation energies for each conformer to further simplify the model. 

During the fitting procedure, we noticed that the intense C=N stretching bands near 1600 cm$^{-1}$ dominate the simVCD metric (see Figures S5 and S6, as well as the discussion in SI). These bands are also overestimated in the calculated IR spectrum, suggesting that their atomic polar tensor (APT) contribution is also too large. Because the APT enters both the MFP-VCD and the enhancement term, the same overestimation is expected in the enhanced VCD spectrum. Optimizing the full spectral range can therefore improve the apparent similarity by fitting these dominant features while giving a less balanced description of the lower-frequency region. To avoid overfitting to these dominant bands, we excluded the C=N-dominated region from the optimization objective, while still reporting the final simVCD over the full spectrum. 

\begin{figure}[h]
    \centering
    \includegraphics[width=9cm]{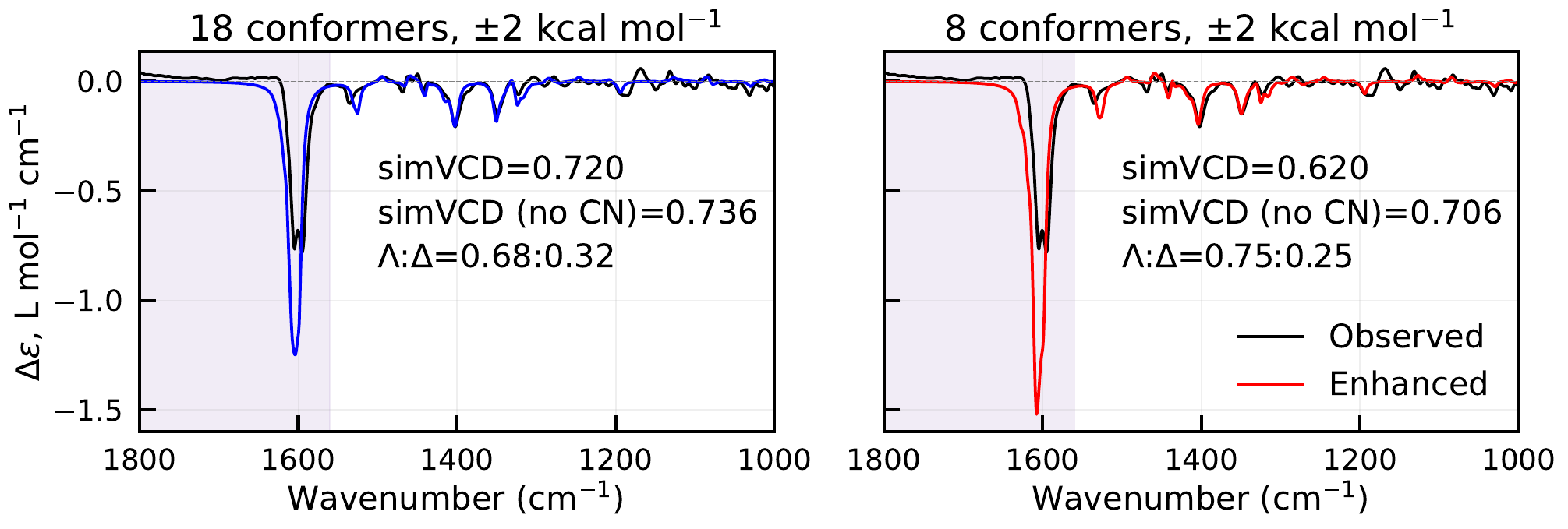}
    \caption{Enhanced VCD spectra averaged over 18 conformers (left) and 8 spectrally unique conformers (right). One excited state was included in the enhancement correction.}
    \label{fig:conf_sets}
\end{figure}

\begin{figure}[!t]
    \centering
    \includegraphics[width=8cm]{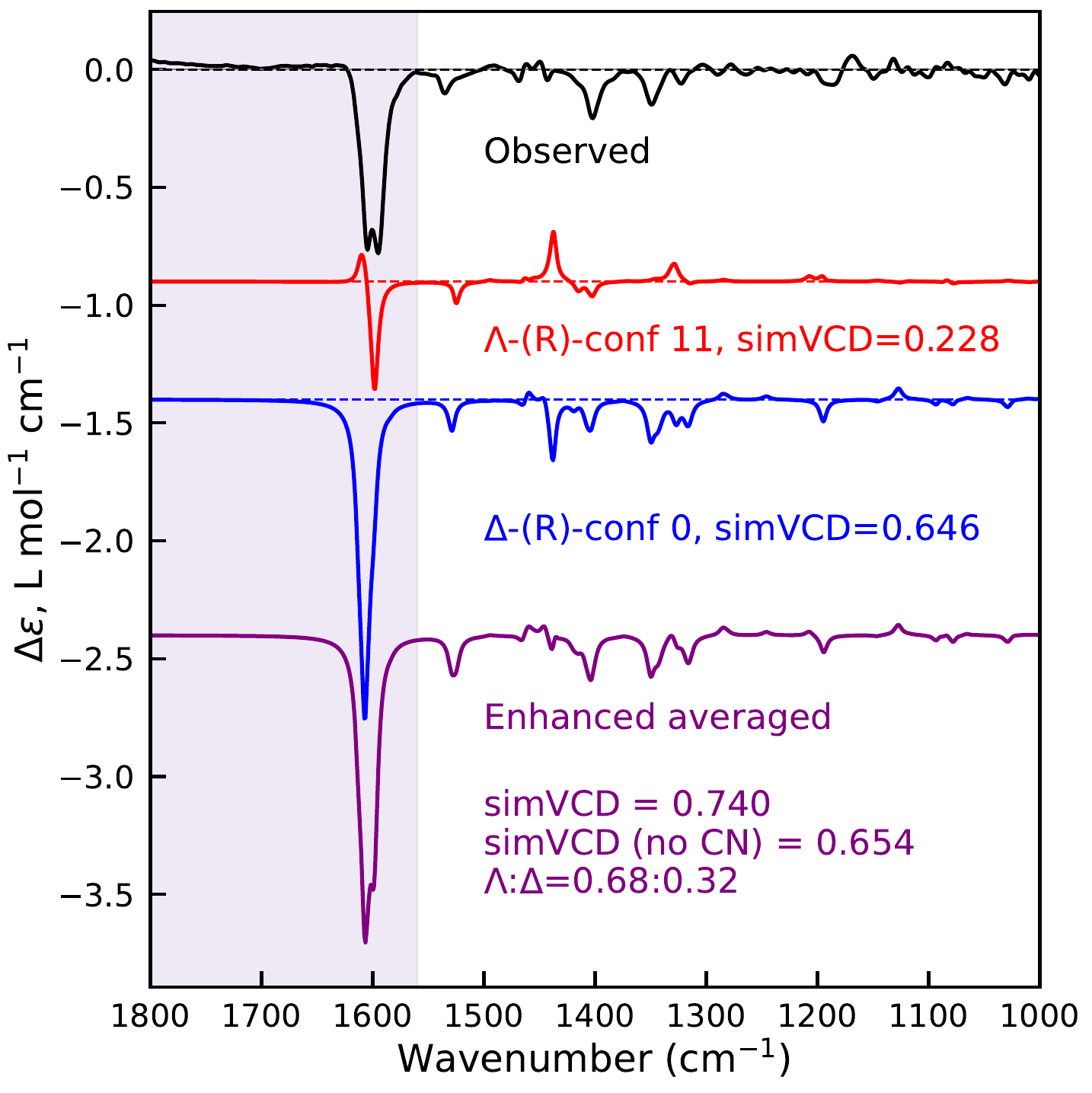}
    \caption{Enhanced VCD spectra averaged over the lowest-energy $\Lambda$ and $\Delta$ conformers (purple) with individual weighted contributions of the conformers $\Lambda-(R)$-conf 11 (blue) and $\Delta-(R)$-conf 0 (red). The fitted excitation energy is 2391 cm$^{-1}$.}
    \label{fig:enhanced_vcd}
\end{figure}

~\autoref{fig:conf_sets} reports enhanced VCD fits using the full 18 conformer set and a reduced set without spectrally redundant conformers. 
 Irrespective of the further details it can be concluded that with our approach we can reproduce very nicely the experimental spectrum, which with other methodologies remained out of reach. 
The simVCD value excluding the CN region (indicated in the plot as no CN) is the optimization objective for Gibbs free and excitation energies, whereas simVCD values are calculated over the full 1800-1000 cm$^{-1}$ range. We used an uncertainty range of $\pm$2 kcal/mol for the Gibbs free energies.
In the SI we demonstrate that increasing the uncertainty from $\pm$1 to $\pm$2 kcal/mol substantially improves the similarity, while a further increase does not significantly affect the results (see Figures S5 and S6).
Excluding the intense C=N bands, the calculated and experimental spectra show a very good agreement with simVCD values above 0.7. However, even with inclusion of these bands still simVCD values are obtained that are quite acceptable. 

Reducing the conformer set by 10 conformers does not qualitatively affect the results, the largest difference occurring in the simVCD value for the entire spectrum which is reduced from 0.72 to 0.62. We attribute this difference mainly to the integrated intensity of the C=N bands, the overall spectrum remains quite similar. As expected, also with respect to other parameters a reduction of the conformer set only leads to minor changes. The estimated $\Lambda$:$\Delta$ ratio does not change significantly (from 0.68:0.32 to 0.75:0.25) while the estimated excitation energy remains more or less the same (2275 cm$^{-1}$ and 2178 cm$^{-1}$ for full and reduced sets with $\pm$2 kcal/mol uncertainties). These results suggest that the smaller conformer set describes the essential enhancement effects.

Finally, we tested what would be the smallest conformer model sufficient to include the enhancement effect. A single $\Lambda$ conformer, $\Lambda$-(R)-conf 11, already reproduces the main negative enhanced VCD features, but the agreement improves substantially when the lowest energy $\Delta$ conformer, $\Delta$-(R)-conf 0, is included as well. The two conformer model gives a spectrum comparable to the larger 18- and 8-conformer models and yields a $\Lambda$:$\Delta$=0.68:0.32 ratio (see ~\autoref{fig:enhanced_vcd}). This indicates that the dominant information in the enhanced VCD spectrum is the balance between $\Lambda$ and $\Delta$ contributions depicted in the same Figure, while the detailed distribution among spectrally similar conformers is not uniquely determined by the VCD data. The fitted excitation energy 2391 cm$^{-1}$ is in good agreement with experimental excitation energy.

\FloatBarrier

\subsection{Approximate symmetry analysis}

\begin{table*}[]
\centering
\caption{Distribution of the first excited state enhancement contribution by approximate $C_2$ mode character for the $\Delta$ and $\Lambda$ conformers. Mean and sum enhancement $|R_i^{\mathrm{enh},1}|$ are given in $10^{-44}$ esu$^2$ cm$^2$. }
\label{tab:approx_c2_enhancement_lambda_delta}
\begin{tabular}{l|ccc|ccc}
\hline
 & \multicolumn{3}{c}{$\Delta$-conf 0} & \multicolumn{3}{c}{$\Lambda$-conf 11} \\
\cline{2-4}\cline{5-7}
Mode class
& $N$ & Mean $|R_i^{\mathrm{enh}}|$ & Sum $|R_i^{\mathrm{enh}}|$
& $N$ & Mean $|R_i^{\mathrm{enh}}|$ & Sum $|R_i^{\mathrm{enh}}|$ \\
\hline
A-like          & 18 & 1.9   & 35.0   & 5  & 2.7   & 13.3 \\
Partly A-like   & 4  & 18.3  & 73.3   & 14 & 35.9  & 502.4 \\
Mixed           & 22 & 29.1  & 640.9  & 29 & 65.7  & 1904.3 \\
Partly B-like   & 4  & 52.7  & 210.9  & 13 & 119.8 & 1557.1 \\
B-like          & 18 & 213.9 & 3849.9 & 5  & 611.6 & 3058.1 \\
\hline
\end{tabular}
\end{table*}

For an ideal $C_2$ complex, it is expected that B electronic transitions are coupled only to B vibrational modes. However, the DFT optimized $\Lambda$-conf 11 and $\Delta$-conf 0 structures are not exactly $C_2$ symmetric. We therefore used an approximate symmetry analysis rather than assigning strict irreducible representations. For each conformer, we determined the best approximate $C_2$ axis. The structure of $\Lambda$-conf 11 shows some deviation from $C_2$ symmetry with a heavy-atom RMSD of 0.262~\AA{} after applying the approximate $C_2$ operation. The core RMSD, which excludes the flexible aryl groups, is substantially smaller (0.143~\AA{}). In contrast, the structure of $\Delta$-conf 0 is much closer to $C_2$ symmetry with heavy-atom and core RMSDs of 0.030 and 0.017~\AA{}, respectively.

\begin{figure*}[]
    \centering
    \begin{subfigure}{8.3cm}
        \centering
        \includegraphics[width=\linewidth]{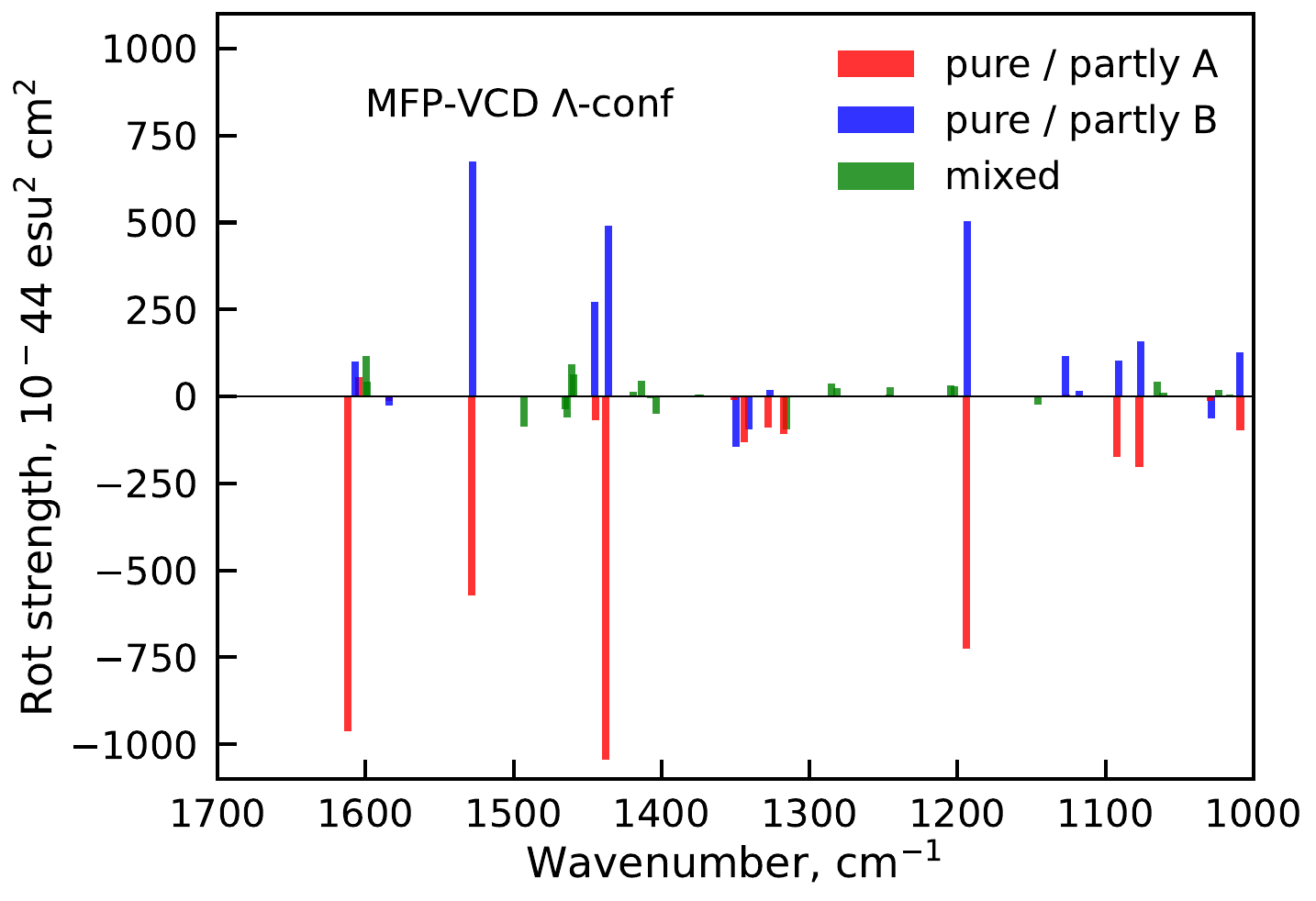}
    \end{subfigure}
    \begin{subfigure}{8.3cm}
        \centering
        \includegraphics[width=\linewidth]{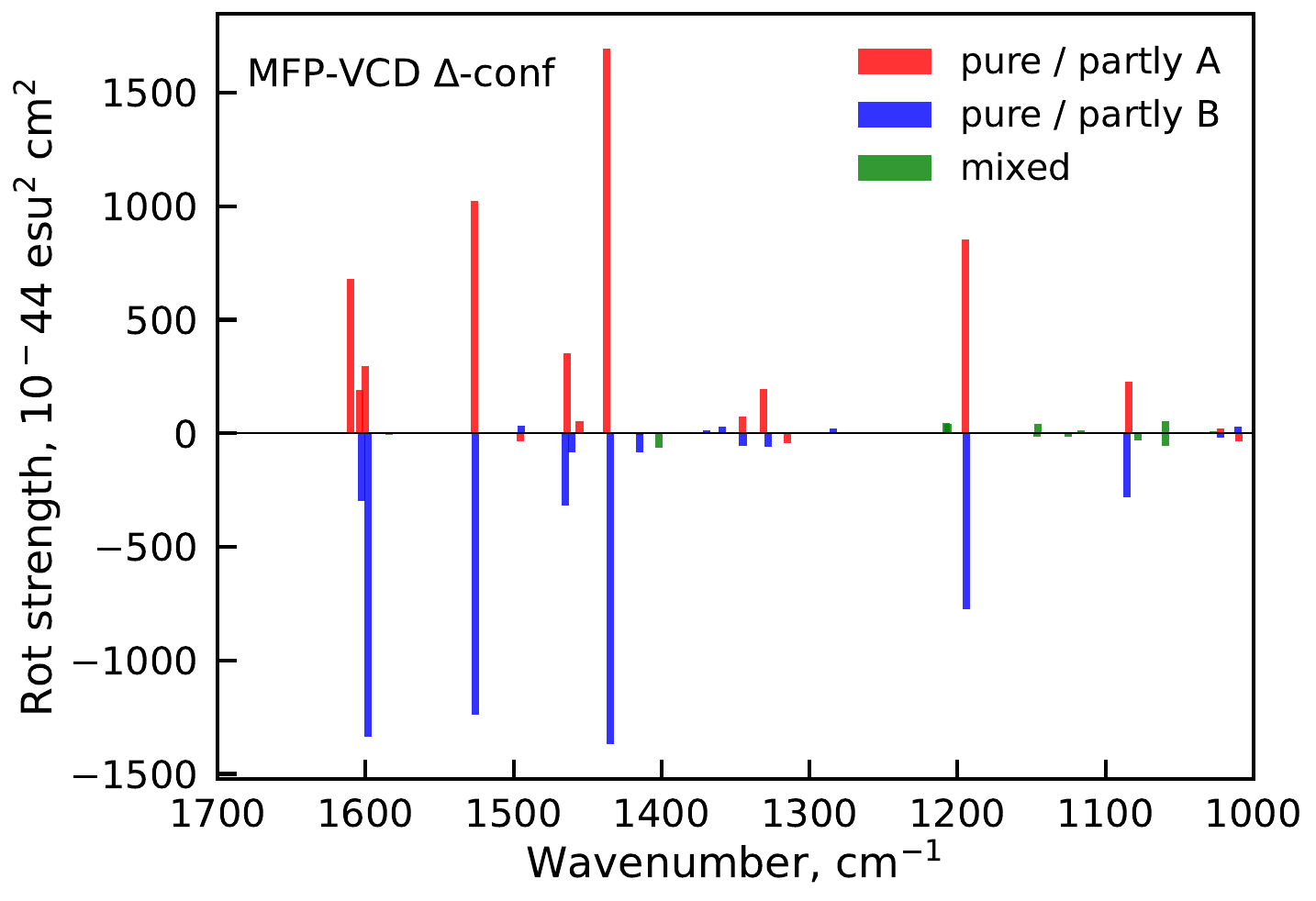}
    \end{subfigure}
    \begin{subfigure}{8.3cm}
        \centering
        \includegraphics[width=\linewidth]{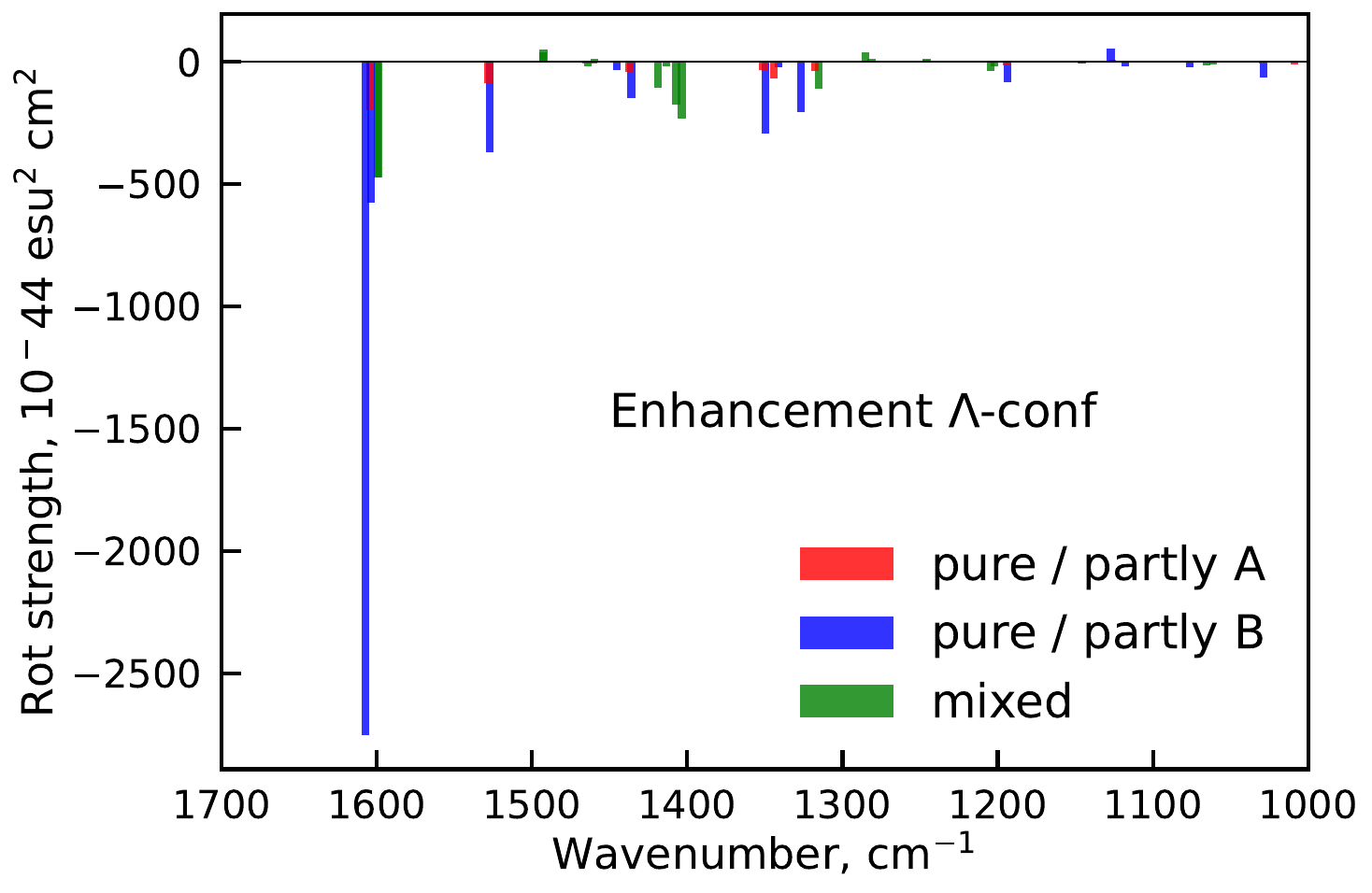}
    \end{subfigure}
    \begin{subfigure}{8.3cm}
        \centering
        \includegraphics[width=\linewidth]{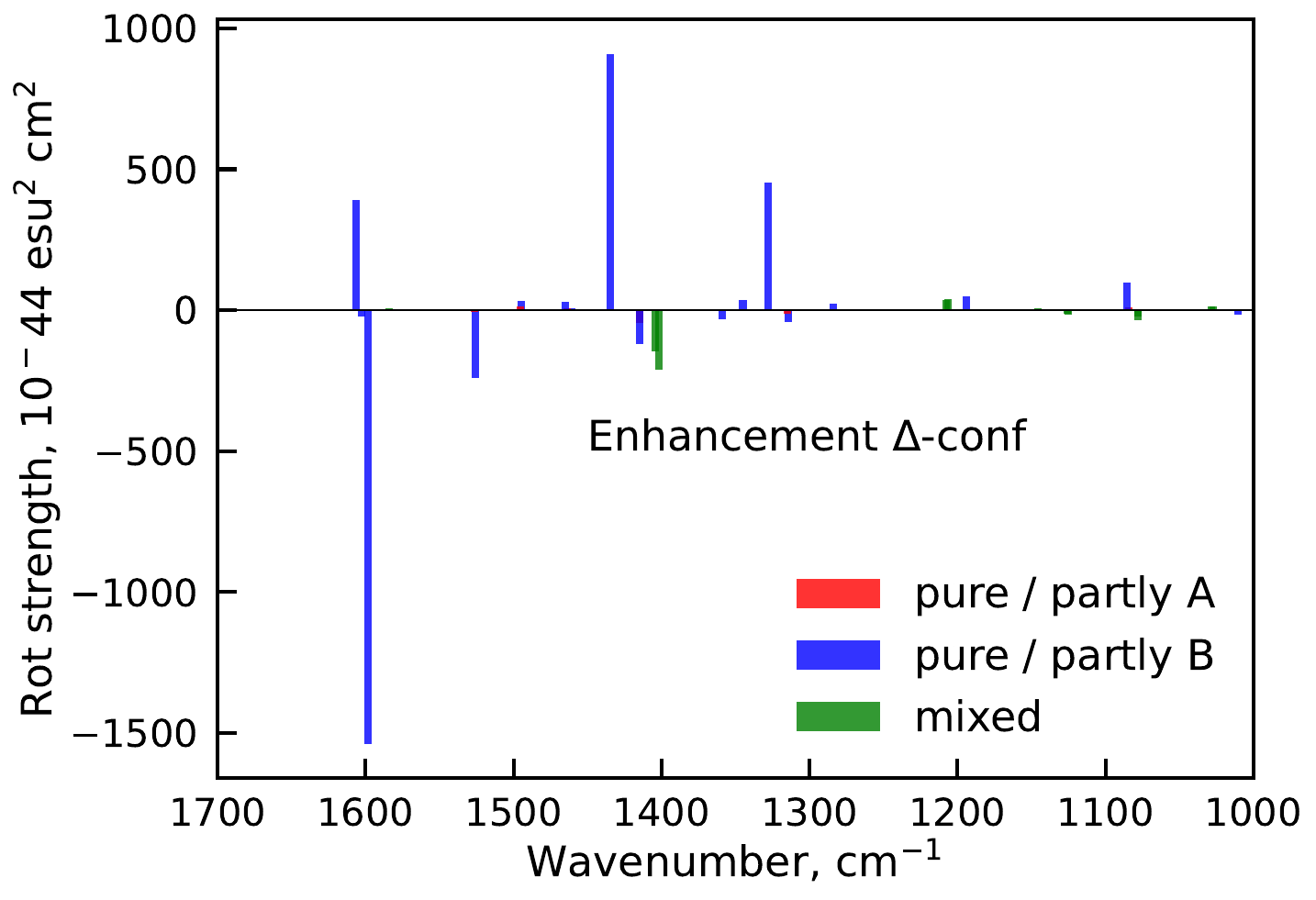}
    \end{subfigure}
    \caption{MFP rotational strengths for the lowest-energy $\Lambda$ and $\Delta$ conformers, colored by the approximate $C_2$ mode character (top). VCD enhancement contribution for the lowest-energy $\Lambda$ and $\Delta$ conformers, colored by the approximate $C_2$ mode character (bottom).}
    \label{fig:c2_plots}
\end{figure*}

We then estimated the symmetry character of each normal mode using a continuous score, $s_i$. This score was calculated as the normalized overlap between the original normal mode displacement and its symmetry-related displacement. Modes with $|s_i|>0.8$ were classified as A-like or B-like, modes with $0.4< |s_i|<0.8$ as partly A-like or partly B-like, and modes with $|s_i| < 0.4$ as mixed. The MFP-VCD rotational strengths reveal a clear dependence on the approximate $C_2$ mode character (see ~\autoref{fig:c2_plots}. For $\Lambda$-conf 11, A-like and partly A-like modes give predominantly negative VCD bands, whereas B-like and partly B-like modes are mostly positive or weakly mixed. We note that this trend is observed mainly in the fingerprint region. Upon changing the metal center chirality from $\Lambda$ to $\Delta$, this MFP sign pattern is reversed as expected for a VCD response. The enhancement contribution from the first low-lying state remains concentrated on B-like modes, which confirms that the enhancement is symmetry-selective even for the approximately $C_2$-symmetric optimized structures. The observed monosignate spectral shape arises primarily from the negative MFP-VCD signal and negative enhancement contributions of the $\Lambda$ structure. However, the enhancement is not uniformly negative: the $\Delta$ conformer exhibits clear positive contributions, which are canceled by the corresponding negative peaks of the $\Lambda$ conformer.

The enhancement contributions for different mode character are summarized in \autoref{tab:approx_c2_enhancement_lambda_delta}. For both conformers, the largest mean enhancement is found for B-like modes, and the enhancement magnitude increases systematically with increasing B character. Thus, the approximate symmetry analysis provides a numerical confirmation of the expected selection rule that the low-lying B-like electronic transition predominantly enhances B-like vibrational modes.

\section{Conclusions}
In this work we have extended our recently introduced enhanced VCD approach to the salicylaldiminato Co(II) Schiff-base complex, a system for which both conformational flexibility and metal center chirality determine the experimental spectrum. DFT calculations find Boltzmann populations that strongly favor the $\Lambda$ form, which is at odds with experimental results that suggest a $\Lambda$:$\Delta$ ratio of about 80:20. In line with our previous studies in which we concluded that calculated Gibbs energies inherently have an uncertainty, we therefore introduced such an uncertainty in the conformer Gibbs free energies. Similarly, an uncertainty was introduced in the calculated excitation energies of electronically excited states that enter the VCD enhancement term. Because the first excited state is well separated from the higher states, the enhancement effects could be described with a single electronic transition. With a $\pm$2 kcal/mol conformer-energy uncertainty and single state enhancement, the calculated VCD spectra reproduce excellently the experimental spectra, reaching simVCD values above 0.7.

An important conclusion coming forward from our studies is that fitting the enhanced VCD spectrum with conformers within a certain energy range does not necessarily lead to a unique determination of the population of every single conformer as different conformers can very well have VCD spectra that for all practical purposes are similar. Fits using the spectra of all conformers thus effectively leads to an overfitting. We have shown that this can be avoided by clustering conformers instead into spectral clusters. We thus found that removing spectrally redundant conformers does not quantitatively change the fitted spectrum and the estimated $\Lambda$:$\Delta$ ratio. In fact, a minimal two-conformer model containing only the lowest-energy $\Lambda$ and $\Delta$ conformers was found to capture very well the main experimental features. This model gives an estimated $\Lambda$:$\Delta$ ratio of approximately 0.68:0.32 and a fitted excitation energy of about 2391 cm$^{-1}$, in good agreement with the experimentally observed NIR CD band.~\cite{Pescitelli2018_schiff}. 

As regarding a further understanding of how bands are enhanced in the VCD spectrum, we have developed and applied an approximate symmetry analysis. Such an analysis predicts that for the case at hand B-like vibrational modes are expected to be enhanced as the electronic transition to the enhancing electronic state has approximately B symmetry. This is indeed what is observed in the experimental VCD spectrum.

Previously, we have developed an approach that enabled us to obtain for rigid systems quantitative agreement between experimentally observed enhanced VCD spectra and theoretically predicted ones. Here, we have demonstrated that this approach remains in full force for conformationally flexible systems. As such, it holds great promise to tackle and elucidate enhanced VCD spectra of larger systems such as metalloproteins. Such studies are presently underway.

\section*{Author Contributions}

\textbf{Mariia Sapova}: Conceptualization; Data Curation; Formal Analysis; Investigation; Methodology; Software; Writing -- original draft; Writing -- review \& editing \\
\textbf{Menno de Boer}: Data Curation; Investigation; Writing -- review \\
\textbf{Wybren J. Buma}: Conceptualization; Funding acquisition; Methodology; Project Administration; Resources;  Supervision; Writing –- review \& editing \\
\textbf{Lucas Visscher}: Conceptualization; Funding acquisition; Methodology; Project Administration; Resources; Supervision; Writing -- review \& editing \\

\section*{Conflicts of interest}

There are no conflicts to declare.

\section*{Data availability}

The data supporting this study, including output files produced by ORCA version 6.0, processed tensor data, experimental spectra, and analysis notebooks required to reproduce the results, are available on GitHub at: \url{https://github.com/LucasVisscher/eVCD}.

\section*{Acknowledgements}\label{sec:acknowledgements}

We acknowledge financial support from NWO via the National Grow Fund Quantum Delta NL (NWO Project Nr. NGF.1582.22.036). We furthermore acknowledge NWO for providing access to Snellius, hosted by SURF through the Computing Time on National Computer Facilities call for proposals.

\input{schiff_complex.bbl}

\bibliographystyle{rsc}
\end{document}


\maketitle
\affiliation
\section{$\Lambda$-(R) and  $\Delta$-(R) conformers}

\begin{table}[htbp]
\centering
\caption{Relative electronic and Gibbs free energies of the $\Lambda$-(R) and $\Delta$-(R) conformers within a 6 kcal/mol energy window. Gibbs weights $w_G$ were calculated at 298.15 K over the full set shown. The numbering of conformers indicates the energetic order predicted by the GOAT~\cite{ORCA_6, GOAT} conformer generation algorithm.}
\label{tab:lambda_delta_energies}
\begin{tabular}{lccccc}
\hline
Conformer & $E$ & $\Delta E$ & $G$  & $\Delta G$ & $w_G$ \\
          &     Eh    & kcal mol$^{-1}$ &   Eh      & kcal mol$^{-1}$ &       \\
\hline
$\Lambda$-(R)-conf~0  & -3721.808690 & 0.639 & -3721.381688 & 0.106 & 0.299 \\
$\Lambda$-(R)-conf~1  & -3721.809702 & 0.004 & -3721.381219 & 0.400 & 0.182 \\
$\Lambda$-(R)-conf~7  & -3721.809090 & 0.388 & -3721.380667 & 0.747 & 0.102 \\
$\Lambda$-(R)-conf~11 & -3721.809709 & 0.000 & -3721.381857 & 0.000 & 0.358 \\
$\Lambda$-(R)-conf~13 & -3721.808228 & 0.929 & -3721.380029 & 1.147 & 0.052 \\
$\Lambda$-(R)-conf~14 & -3721.803403 & 3.957 & -3721.375731 & 3.844 & 0.001 \\
$\Lambda$-(R)-conf~15 & -3721.800655 & 5.681 & -3721.373415 & 5.298 & 0.000 \\
$\Lambda$-(R)-conf~16 & -3721.800458 & 5.805 & -3721.373198 & 5.433 & 0.000 \\
\hline
$\Delta$-(R)-conf~0  & -3721.805905 & 2.387 & -3721.377837 & 2.522 & 0.005 \\
$\Delta$-(R)-conf~6  & -3721.802550 & 4.492 & -3721.375018 & 4.292 & 0.000\\
$\Delta$-(R)-conf~8  & -3721.803521 & 3.883 & -3721.375195 & 4.180 & 0.000 \\
$\Delta$-(R)-conf~19 & -3721.801657 & 5.053 & -3721.374848 & 4.398 & 0.000 \\
$\Delta$-(R)-conf~29 & -3721.801857 & 4.927 & -3721.373731 & 5.099 & 0.000 \\
$\Delta$-(R)-conf~30 & -3721.802388 & 4.593 & -3721.374830 & 4.410 & 0.000 \\
$\Delta$-(R)-conf~31 & -3721.800278 & 5.918 & -3721.372963 & 5.581 & 0.000 \\
$\Delta$-(R)-conf~36 & -3721.802609 & 4.455 & -3721.373863 & 5.016 & 0.000 \\
   
$\Delta$-(R)-conf~38 & -3721.803113 & 4.139 & -3721.374281 & 4.754 & 0.000 \\
$\Delta$-(R)-conf~44 & -3721.801670 & 5.044 & -3721.373280 & 5.382 & 0.000 \\
\hline
\end{tabular}
\end{table}

\begin{table}[htbp]
\centering
\caption{Vertical excitation energies (cm$^{-1}$) from TD-B3LYP~\cite{becke1993density} and CASSCF(7,5)/NEVPT2 calculations.}
\label{tab:vertical_excitations_cm}
\small
\setlength{\tabcolsep}{4pt}
\begin{tabular}{ccccccccc}
\hline
\multicolumn{9}{c}{TD-B3LYP, $\Lambda$ conformers} \\
\hline
\# & $\Lambda$-conf 0 & $\Lambda$-conf 1 & $\Lambda$-conf 7 & $\Lambda$-conf 11 & $\Lambda$-conf 13 & $\Lambda$-conf 14 & $\Lambda$-conf 15 & $\Lambda$-conf 16 \\
\hline
1 & 6340 & 6364 & 6622 & 6202 & 6509 & 6509 & 6412 & 6452 \\
2 & 9913 & 9784 & 9590 & 9703 & 9961 & 9719 & 9477 & 9429 \\
3 & 10308 & 10251 & 10816 & 10171 & 11090 & 10396 & 10493 & 10679 \\
4 & 11977 & 12098 & 11937 & 12227 & 11743 & 12518 & 12582 & 12179 \\
5 & 16131 & 15913 & 17389 & 15712 & 17623 & 15841 & 16228 & 16567 \\
6 & 18873 & 19027 & 18123 & 19059 & 17833 & 19422 & 19511 & 19075 \\
\hline
\multicolumn{9}{c}{TD-B3LYP, $\Delta$ conformers} \\
\hline
\# & $\Delta$-conf 0 & $\Delta$-conf 6 & $\Delta$-conf 8 & $\Delta$-conf 19 & $\Delta$-conf 29 & $\Delta$-conf 30 & $\Delta$-conf 31 & $\Delta$-conf 36 \\
\hline
1 & 6646 & 6743 & 6348 & 6493 & 6356 & 6331 & 6638 & 6235 \\
2 & 10477 & 9945 & 9767 & 10380 & 9429 & 9775 & 10138 & 8824 \\
3 & 11155 & 11155 & 10251 & 10711 & 10655 & 10606 & 11268 & 10179 \\
4 & 11477 & 11493 & 12671 & 11872 & 11832 & 12171 & 11469 & 12687 \\
5 & 17293 & 17438 & 15913 & 16865 & 16058 & 16252 & 17696 & 14849 \\
6 & 17728 & 17881 & 19252 & 18349 & 18970 & 19115 & 17825 & 19994 \\
\hline
\# & $\Delta$-conf 38 & $\Delta$-conf 44 & & & & & & \\
\hline
1 & 6275 & 6017 & & & & & & \\
2 & 9082 & 8727 & & & & & & \\
3 & 10235 & 10001 & & & & & & \\
4 & 12873 & 13413 & & & & & & \\
5 & 15187 & 14857 & & & & & & \\
6 & 19986 & 20712 & & & & & & \\
\hline
\multicolumn{9}{c}{CASSCF/NEVPT2} \\
\hline
\# & $\Lambda$-conf 11 & $\Delta$-conf 0 & & & & & & \\
\hline
1 & 2354 & 2602 & & & & & & \\
2 & 6068 & 7397 & & & & & & \\
3 & 8303 & 7742 & & & & & & \\
4 & 8784 & 9440 & & & & & & \\
5 & 9519 & 9670 & & & & & & \\
6 & 10053 & 10098 & & & & & & \\
\hline
\end{tabular}
\end{table}

\begin{figure}[ht]
    \centering
    \includegraphics[width=0.9\textwidth]{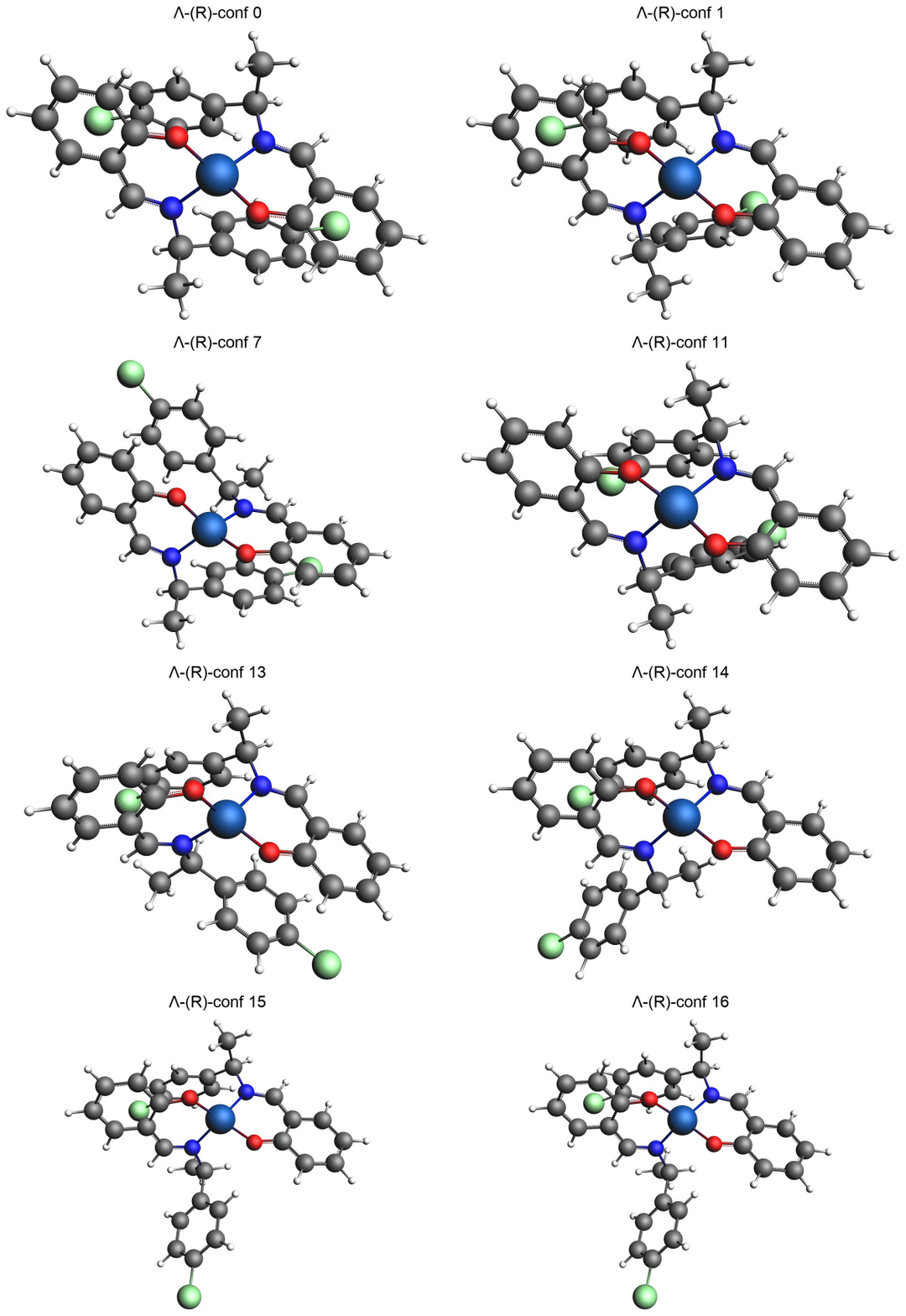}
    \caption{$\Lambda$ conformer structures within 6 kcal/mol energy window. $\Lambda$-(R)-conf 11 is the lowest lying conformer.}
    \label{fig:lambda_confs}
\end{figure}

\begin{figure}[ht]
    \centering
    \includegraphics[width=0.9\textwidth]{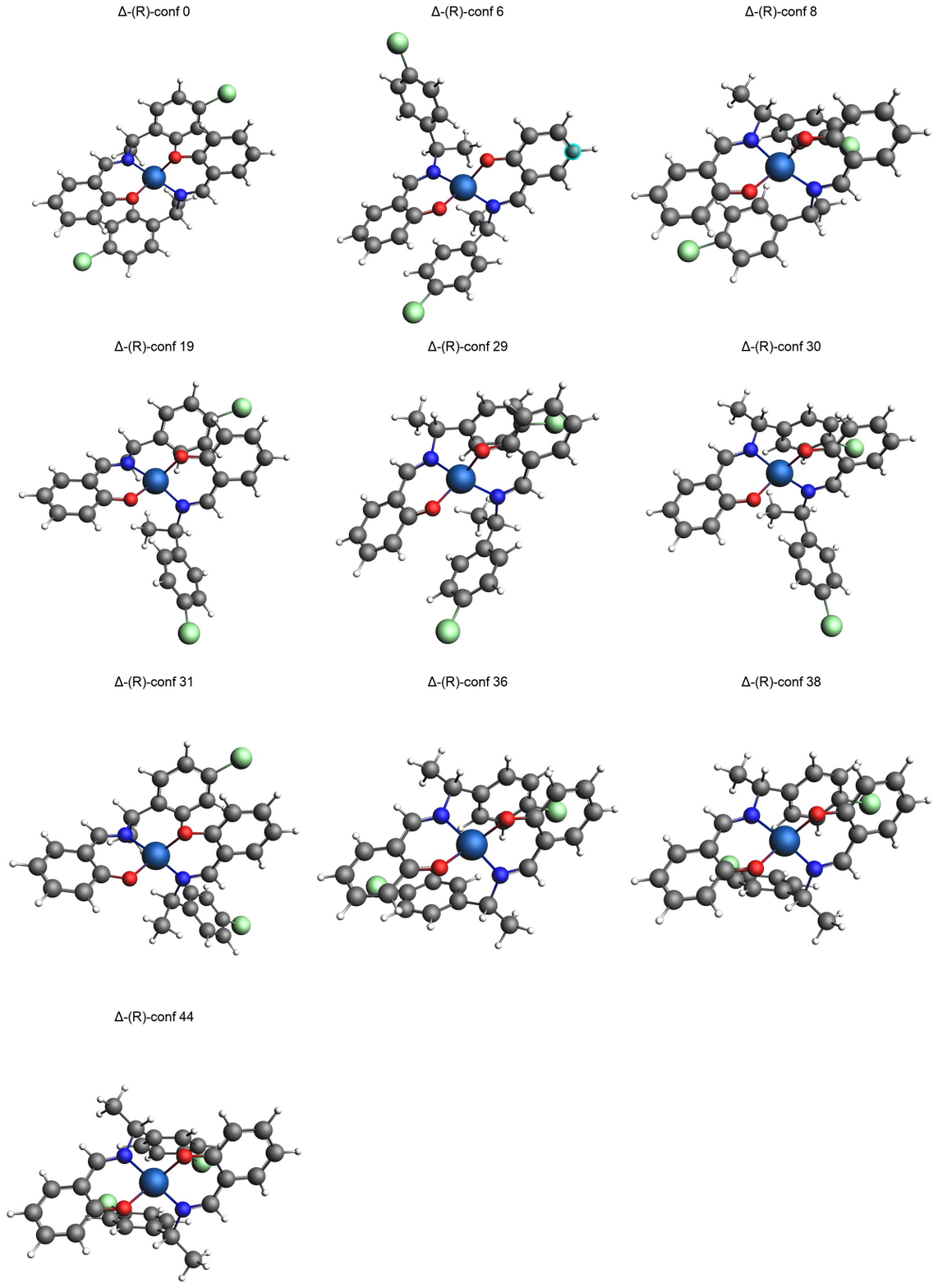}
    \caption{$\Delta$ conformer structures within 6 kcal/mol energy window.}
    \label{fig:delta_confs}
\end{figure}

\begin{figure*}[h]
    \centering
    \includegraphics[width=0.75\textwidth]{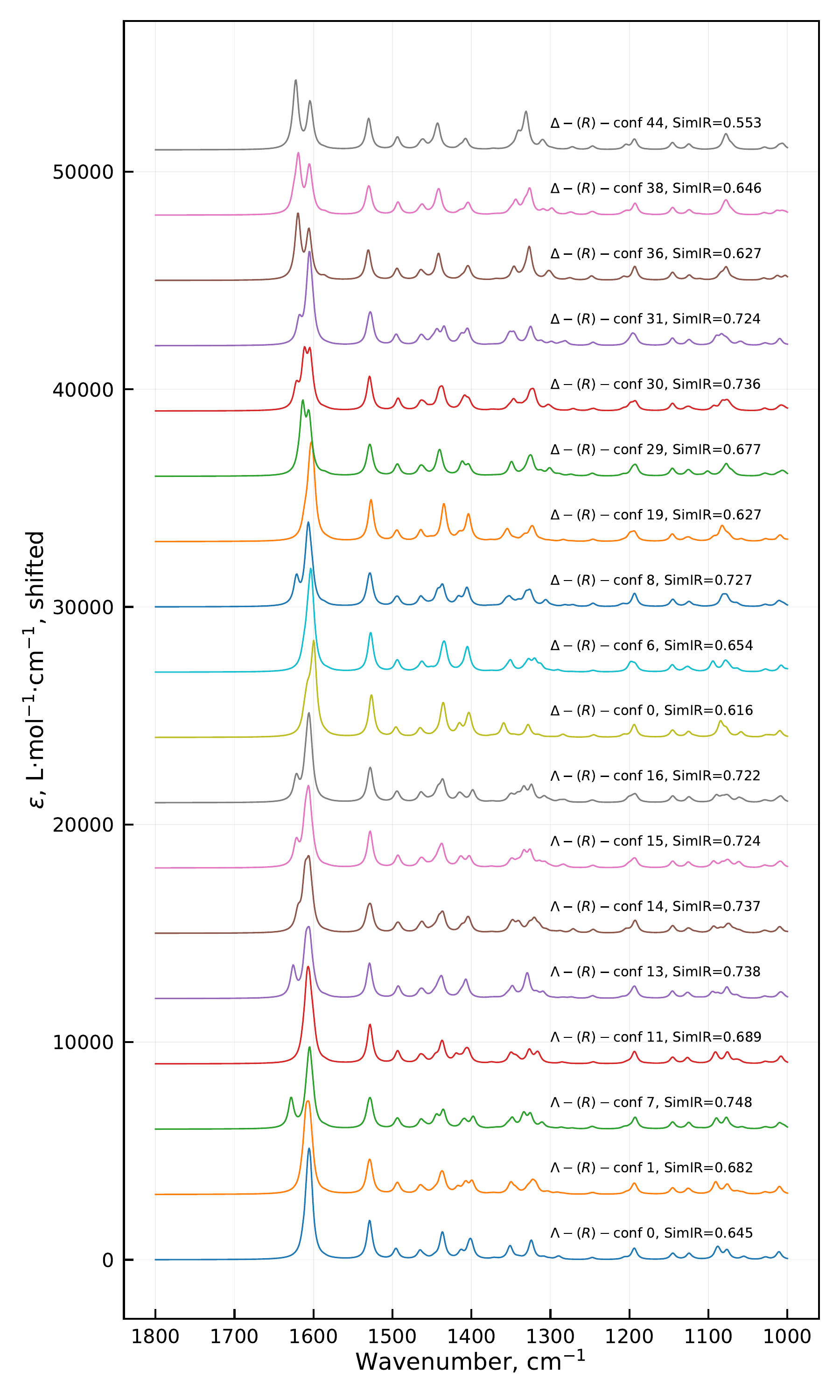}
    \caption{IR spectra of unique conformers calculated with B3LYP-D3BJ/def2-TZVP/CPCM.}
    \label{fig:ir_conf_b3lyp}
\end{figure*}

\begin{figure*}[h]
    \centering
    \includegraphics[width=0.75\textwidth]{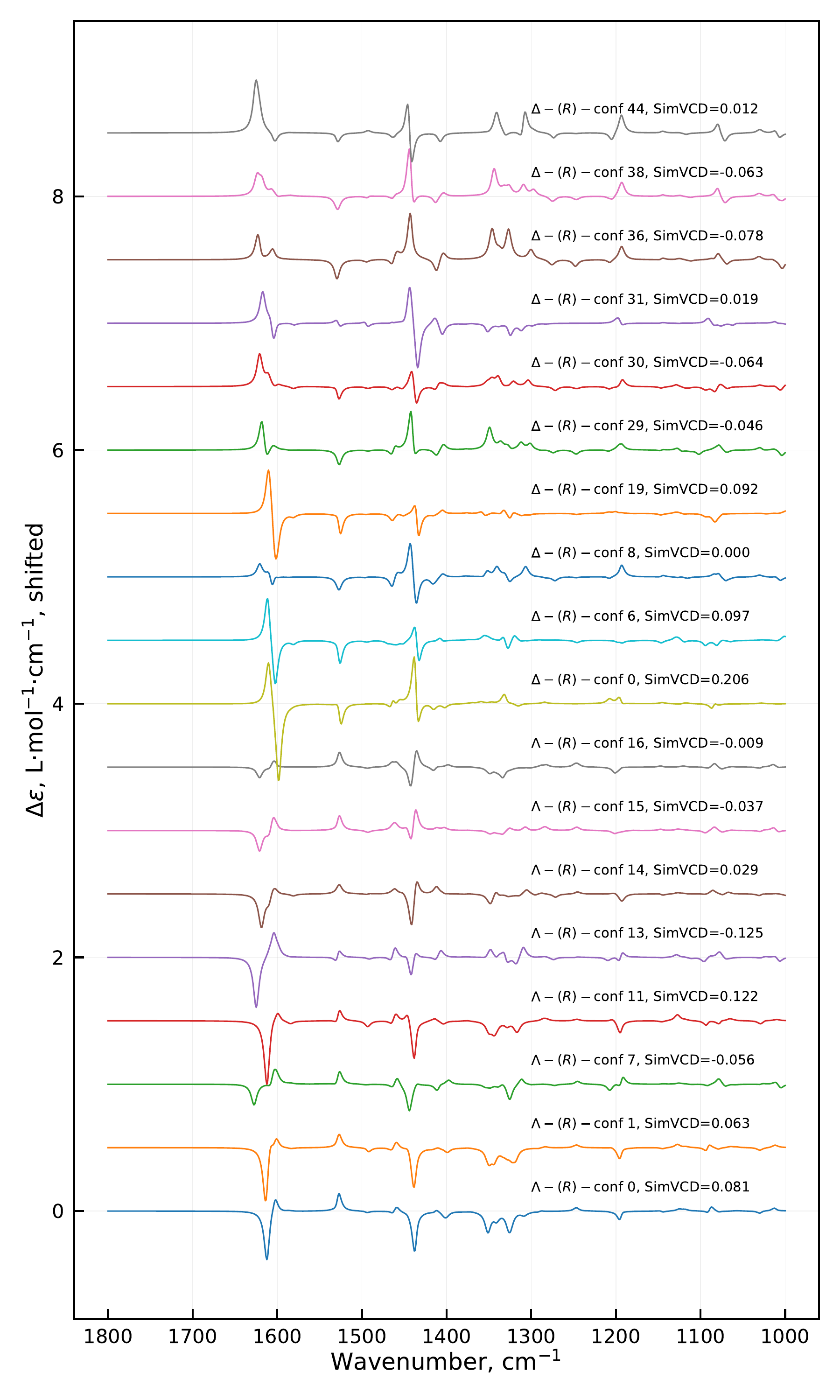}
    \caption{MFP-VCD spectra of unique conformers calculated with B3LYP-D3BJ/def2-TZVP/CPCM.}
    \label{fig:vcd_conf_b3lyp}
\end{figure*}

\FloatBarrier

\section{Enhanced VCD calculation}

In the enhanced VCD calculations we include coupling only to the first electronically excited state. This choice is supported by a significant gap between the first and higher electronically  excited states obtained both at the TDDFT and CASSCF/NEVPT2 level (Table S2). The second and third excited states, in contrast, are close in energy and change order for some conformers. Because our model assigns the same excitation energies to all conformers, the same electronic state must be followed consistently across the conformer ensemble. In the present system, these higher states make only a minor contribution to the enhancement, so the first excited state provides a compact and quantitative description of the effect.

We then compared enhanced VCD fits obtained with the full conformer set and with a reduced set in which spectrally redundant conformers were removed. In both cases, the fit parameters were the conformer energy shifts within a chosen uncertainty range and a single shared excitation energy. We tested conformer energy uncertainty ranges of $\pm$1, $\pm$2, and $\pm$3 kcal mol$^{-1}$ (Figure S5). Increasing the uncertainty from $\pm$2 to $\pm$3 kcal mol$^{-1}$ was found to only have a minor effect on the optimized spectra and the $\Lambda$:$\Delta$ ratio (see simVCD and $\Lambda$:$\Delta$ in Figure S5). A potential concern is that the intense C=N stretching region near 1600 cm$^{-1}$ dominates the SimVCD metric. This band is also overestimated in the calculated IR spectrum, suggesting that the corresponding VCD intensity may be overestimated as well. Therefore, fitting the full spectral range can improve the apparent similarity by reproducing this dominant feature while giving less reliable agreement in the lower-frequency region. To test this effect, we evaluated the spectral similarity both over the full range and with the dominant C=N feature excluded. For the reduced model, the full-range SimVCD increases, but the similarity outside the C=N region does not increase correspondingly. This indicates that this improvement is driven mainly by the intense C=N band rather than by a uniformly better description of the spectrum. Therefore, we excluded the C=N-dominated region from the fit while still reporting the final similarity over the full spectral range (Figure S6).
With this more balanced fitting procedure, the full and reduced conformer sets give very similar results. The difference in SimVCD is only about 0.03, and visual inspection concludes that removing 10 spectrally redundant conformers does not qualitatively change the spectrum. The fitted $\Lambda$:$\Delta$ ratio also becomes close to 70:30, in better agreement with the experimental ratio of approximately 80:20.

To investigate whether an even smaller conformer set is sufficient to describe the enhanced VCD spectrum, we fitted the excitation energy for a single $\Lambda$-conf 11 excluding C=N dominated region (Figure S7). This single-conformer model reproduces the main negative features of the experimental spectrum already reasonably well and leads to a fitted excitation energy of 2499 cm$^{-1}$. Including a lowest-energy $\Delta$ conformer ($\Delta$-conf 0) markedly improves the spectra similarity from 0.565 for the single-conformer model to 0.740 for two-conformer model and an excitation energy of 2391.8 cm$^{-1}$. This result remains consistent with what is experimentally observed in terms of transitions to low-lying electronically excited states. Finally, we compared the $\Lambda$:$\Delta$ ratio for three considered conformer sets: (i) 18 conformers within a 6 kcal/mol energy window, (ii) 8 spectrally independent conformers and (iii) the minimal two-conformer set. The resulting ratios are only slightly different: 0.69:0.31, 0.73:0.27 and 0.68:0.32. Thus, the enhanced VCD spectrum is primarily sensitive to the overall $\Lambda$/$\Delta$ balance rather than to the detailed distribution among individual conformers.

\begin{figure*}[h]
    \centering
    \includegraphics[width=0.85\textwidth]{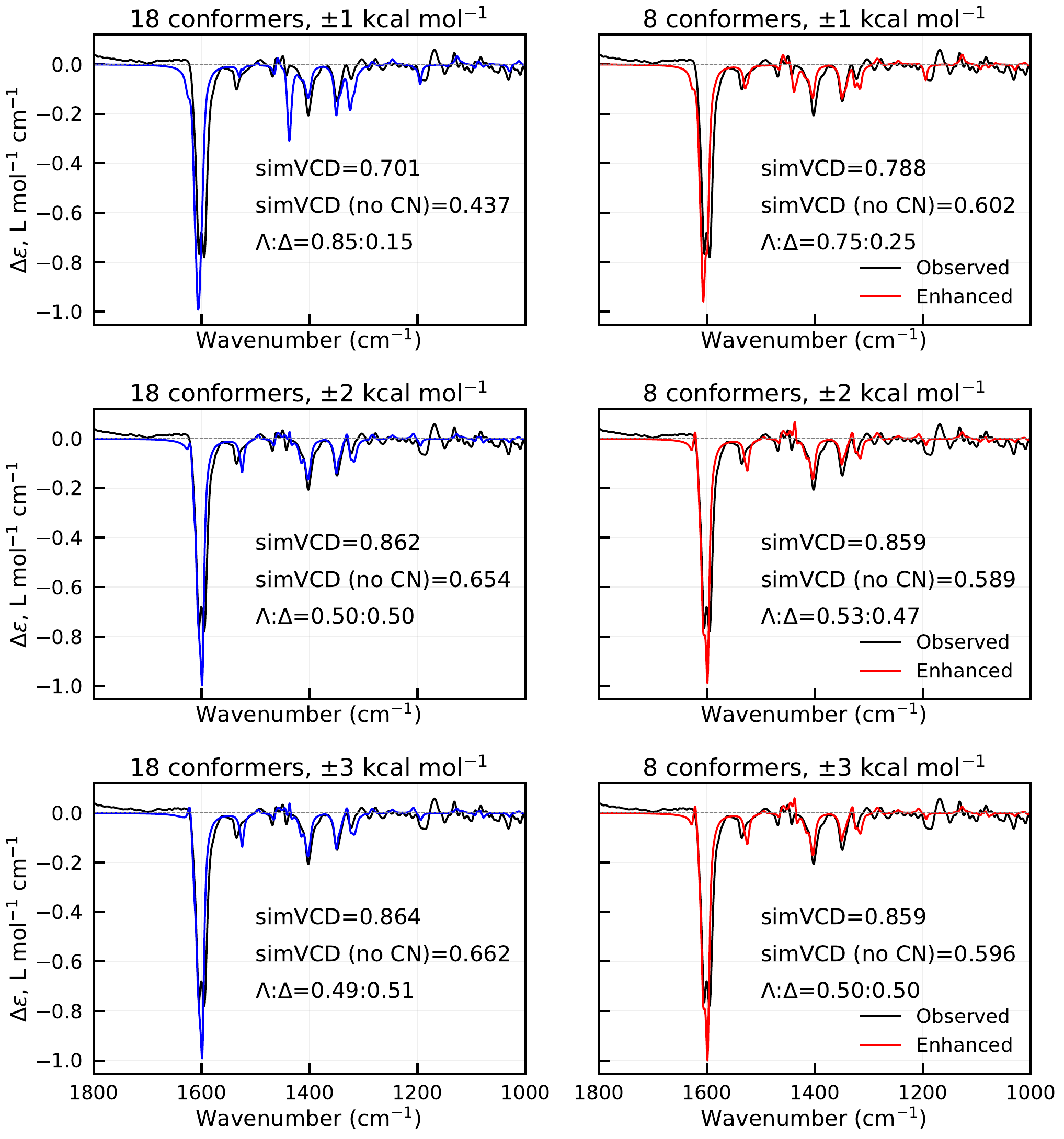}
    \caption{Enhanced VCD spectra calculated with all unique conformers within a 6 kcal/mol energy window (left) and with spectrally unique conformers (right) fitted the full 1000-1800 cm$^{-1}$ region. SimVCD values are calculated on the full frequency range, simVCD (no CN) excludes dominating C=N peaks.}
    \label{fig:ir_conf_b3lyp}
\end{figure*}

\begin{figure*}[h]
    \centering
    \includegraphics[width=0.85\textwidth]{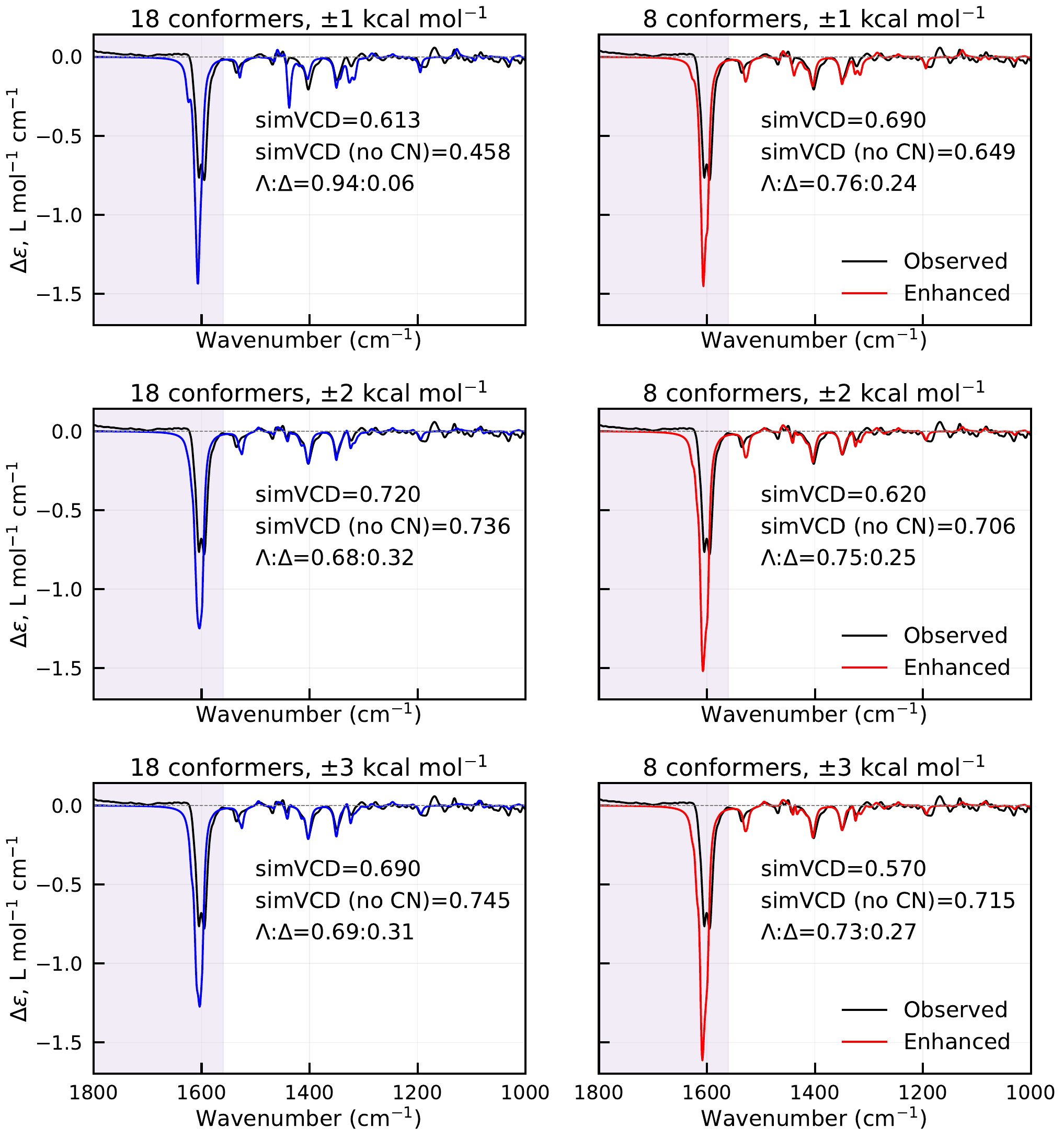}
    \caption{Enhanced VCD spectra calculated with all unique conformers within a 6 kcal/mol energy window (left) and with spectrally unique conformers (right) resulting from fits in which the purple range has been excluded.}
    \label{fig:ir_conf_b3lyp}
\end{figure*}

\FloatBarrier

\begin{figure}[htbp]
    \centering
    \begin{subfigure}[t]{0.49\textwidth}
        \centering
        \includegraphics[width=\linewidth]{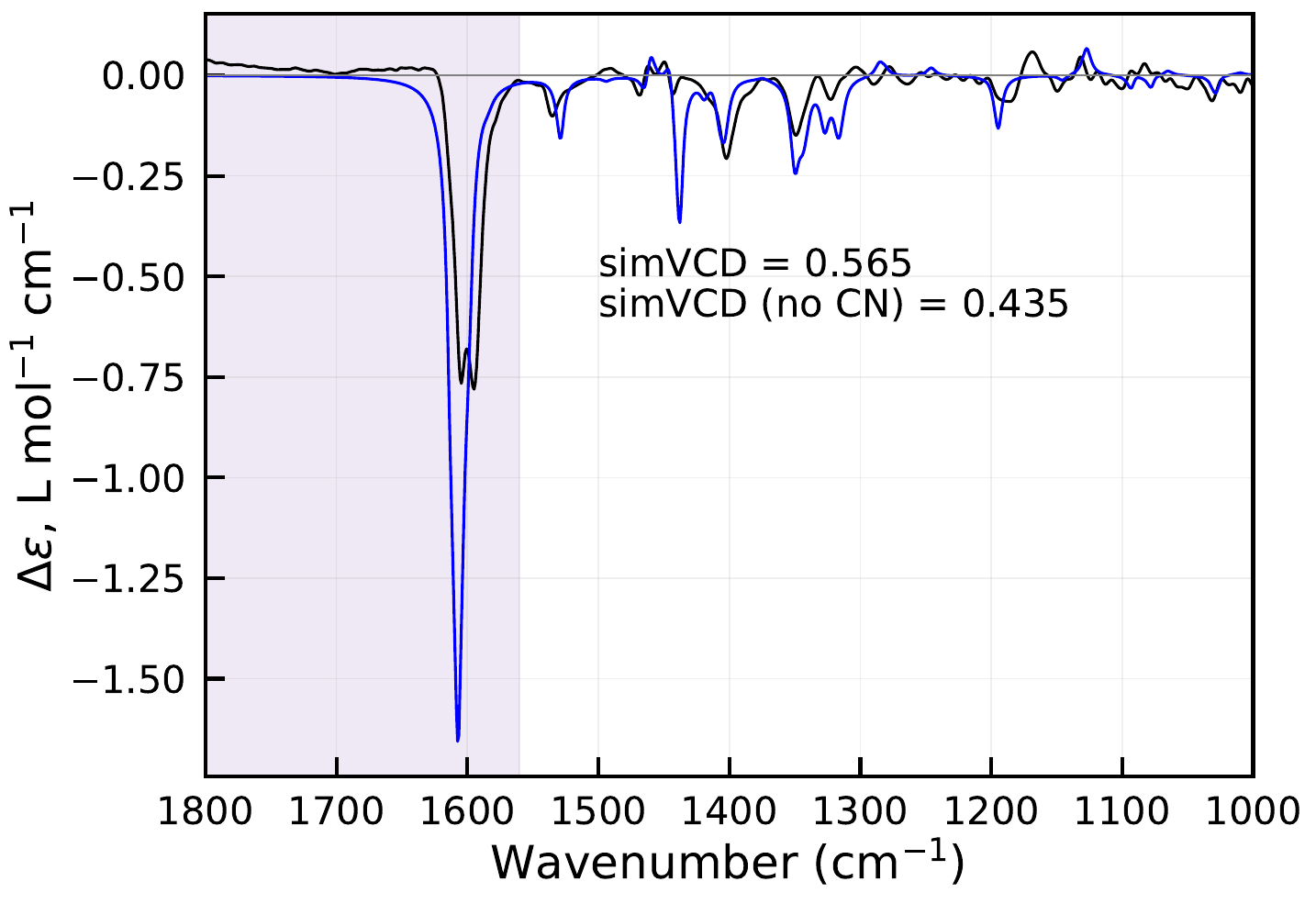}
    \end{subfigure}
    \hfill
    \begin{subfigure}[t]{0.49\textwidth}
        \centering
        \includegraphics[width=\linewidth]{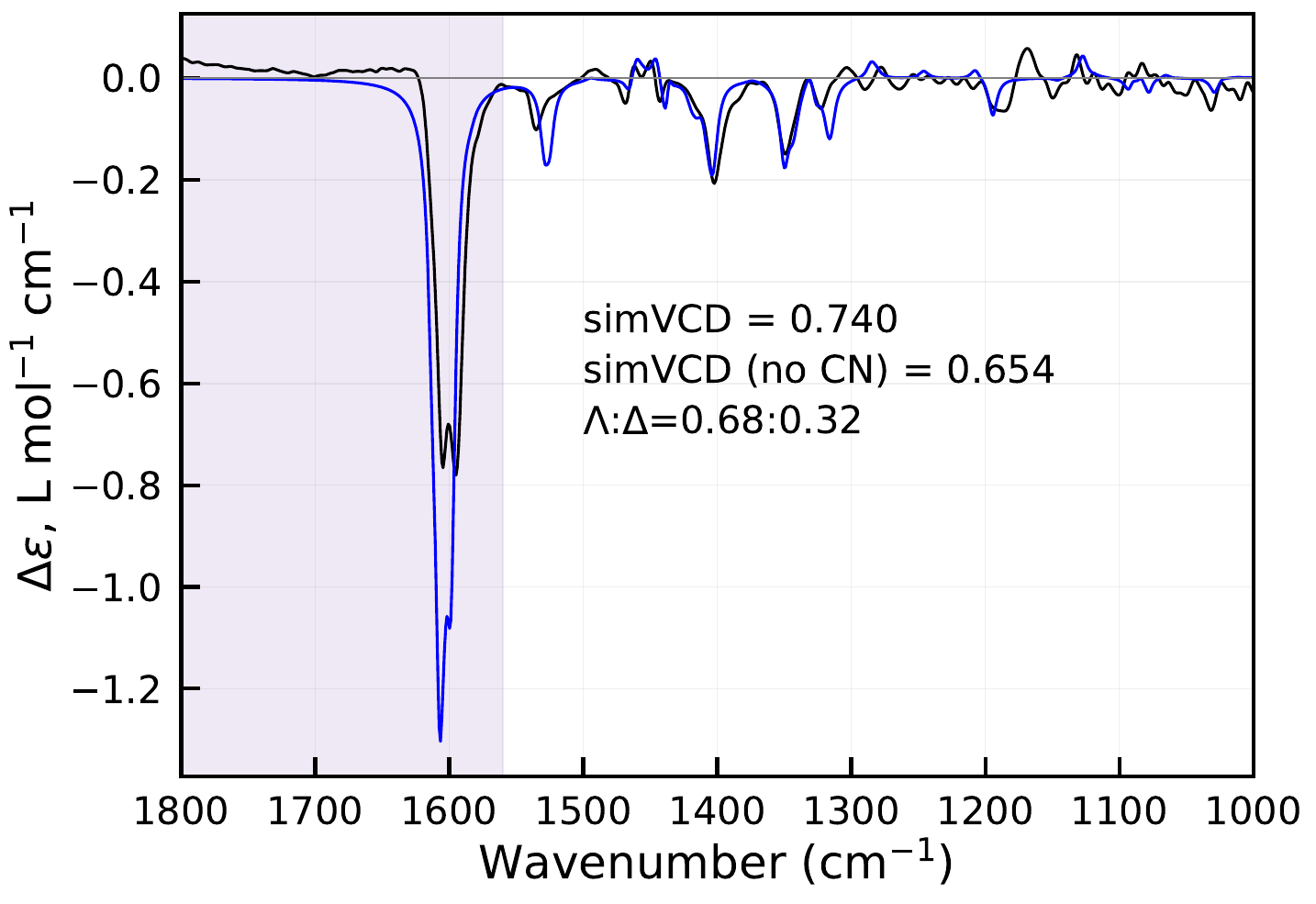}
    \end{subfigure}

    \caption{Enhanced VCD spectra calculated with single conformer $\Lambda$-conf 11 (left), and with two conformers ($\Lambda$-conf 11 and $\Delta$-conf 0) (right).}
    \label{fig:enhanced_vcd_comparison}
\end{figure}

\subsection{Calculations without dispersion correction and solvent model}

We performed additional B3LYP~\cite{becke1993density}/def2-TZVP~\cite{Weigend2005Def2} calculations to reproduce the level of theory used in~\cite{Pescitelli2019}. At this level we reoptimized 26 $\Lambda$ and 51 $\Delta$ structures generated by GOAT~\cite{ORCA_6, GOAT}. Consistent with previous observations~\cite{Pescitelli2018_schiff}, the lowest-energy structure has $\Delta$ metal center chirality, contrary to the experimental preference for the $\Lambda$ form. However, our DFT reoptimization yielded a larger set of unique structures: 9 $\Lambda$ and 11 $\Delta$ conformers within 5 kcal mol$^{-1}$. All of these structures were included in the enhanced VCD calculation. The agreement with experiment is poorer than that for the conformer set optimized with dispersion correction and the solvent model. Nevertheless, allowing uncertainty in the Gibbs free energies ($\pm 3$ kcal mol$^{-1}$) increases the $\Lambda$ population, opposite to the raw DFT Boltzmann weights and leads to qualitative agreement with the experimental data. The optimized model gives a $\Lambda$:$\Delta$ ratio of 63:37 and an excitation energy of 2203 cm$^{-1}$, with SimVCD = 0.595 for the full range and SimVCD = 0.566 when the C=N region is excluded (see Figure S8).

\begin{figure*}[h]
    \centering
    \includegraphics[width=0.5\textwidth]{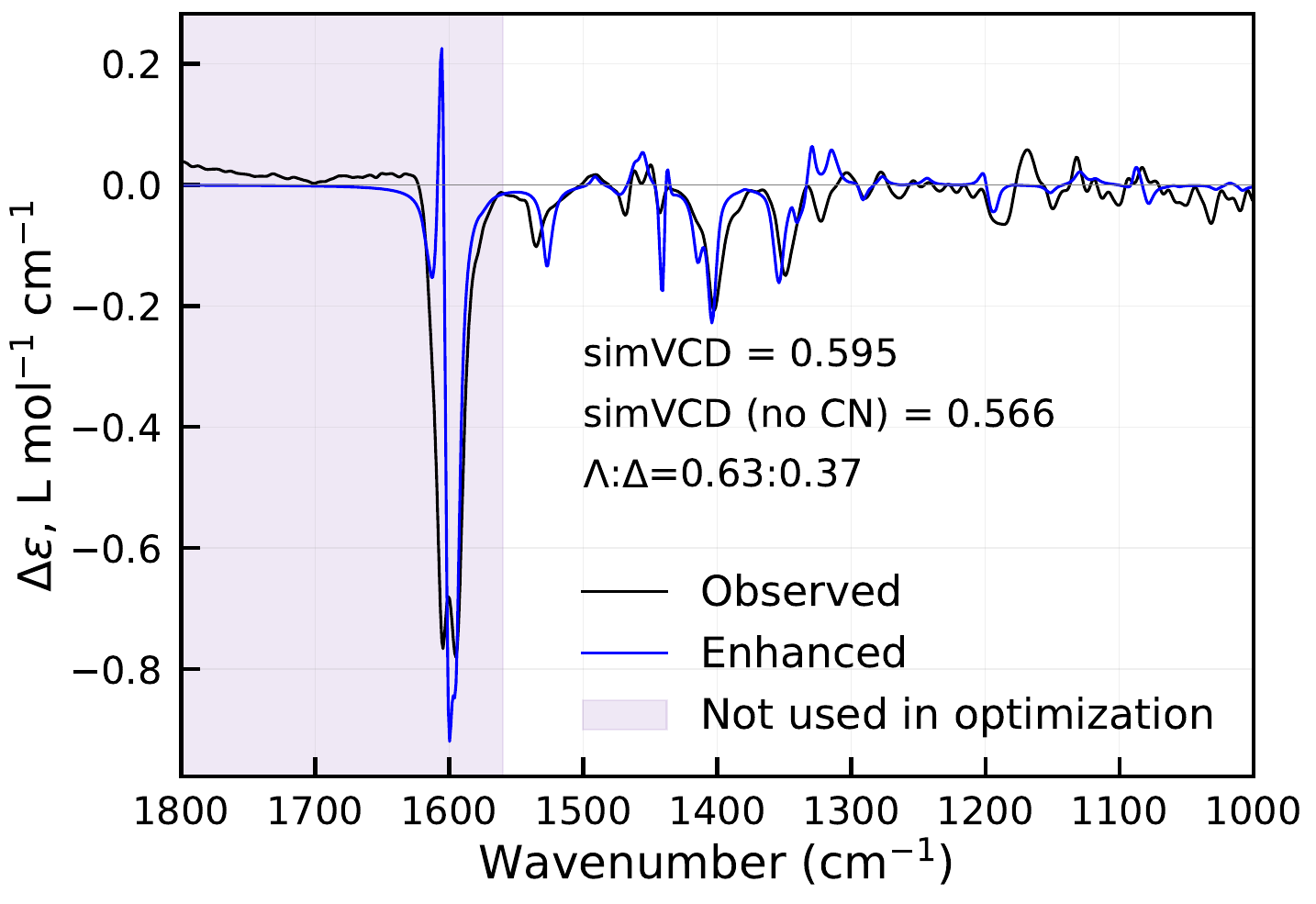}
    \caption{Enhanced VCD spectra calculated with all unique conformers optimized on B3LYP/def2-TZVP level.}
    \label{fig:ir_conf_b3lyp_nosolv}
\end{figure*}

\FloatBarrier

\subsection{Other aryl groups}

For other substituents (X = H, OCH$_3$,Br) we used the two-conformer model and the B3LYP-D3BJ/def2-TZVP/CPCM level of theory. We reoptimized conformers $\Lambda$-conf 11 and $\Delta$-conf 0 conformers. Figure S9 to S11 show that the two considered conformers describe the enhanced VCD spectra well, although separately optimized conformer sets are probably needed for more reliable estimates of the excitation energy and $\Lambda$ population distribution. Similarity values over the full ranges are 0.674 for H, 0.817 for the OCH$_3$, and 0.754 for the Br substituent.

\begin{figure*}[h]
    \centering
    \includegraphics[width=0.95\textwidth]{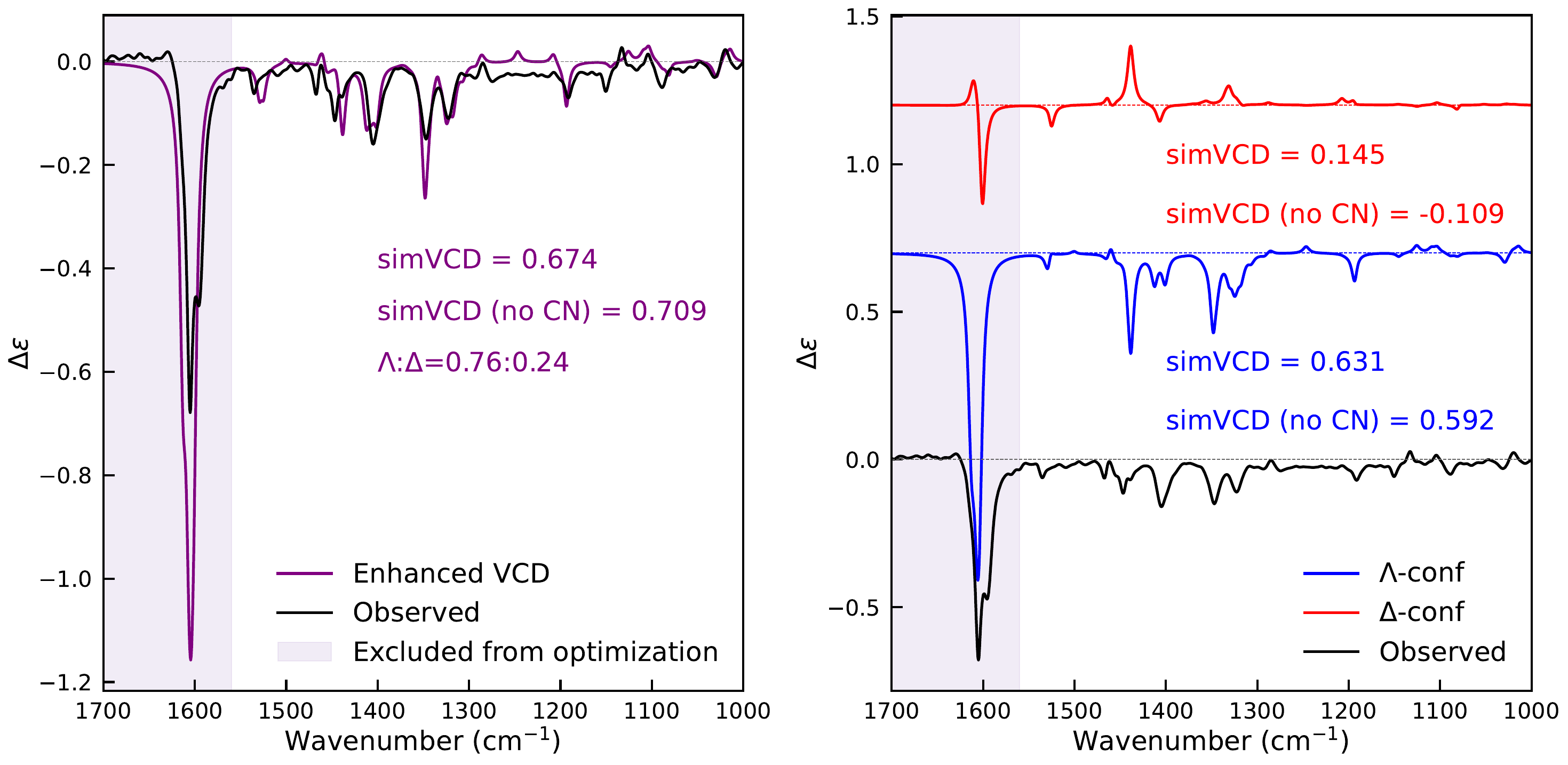}
    \caption{Enhanced VCD spectra of averaged over two $\Lambda$ and $\Delta$ conformers (left) for H as substituent. Individual weighted contributions of the two conformer with the SimVCD values (right). The fitted excitation energy is 2516 cm$^{-1}$.}
    \label{fig:h_subs}
\end{figure*}

\begin{figure*}[h]
    \centering
    \includegraphics[width=0.95\textwidth]{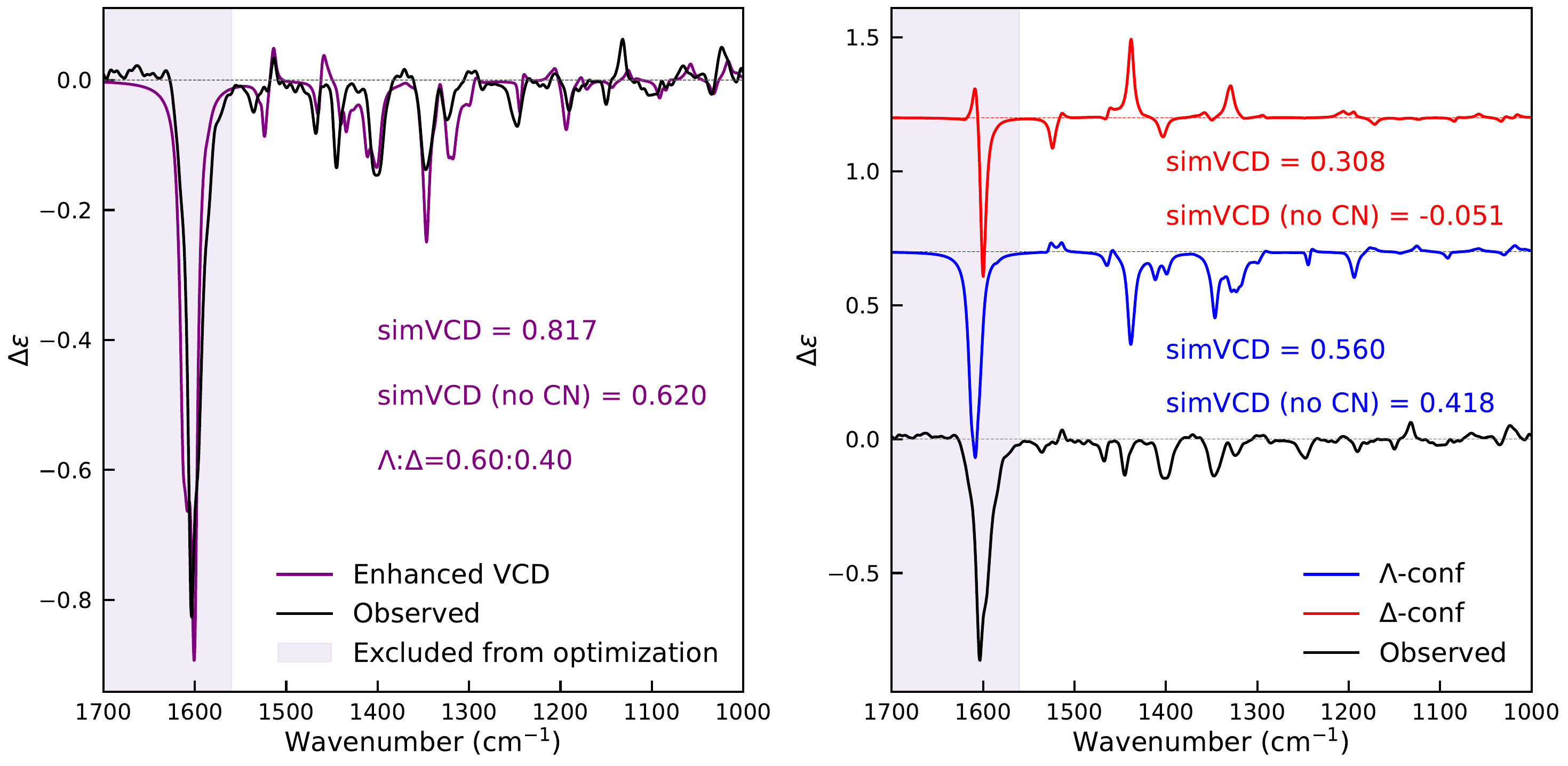}
    \caption{Enhanced VCD spectra averaged over two $\Lambda$ and $\Delta$ conformers (left) for OCH$_3$ as substituent.  Individual weighted contributions of the two conformer with the SimVCD values (right). The fitted excitation energy is 2588 cm$^{-1}$.}
    \label{fig:h_subs}
\end{figure*}

\begin{figure*}[h]
    \centering
    \includegraphics[width=0.95\textwidth]{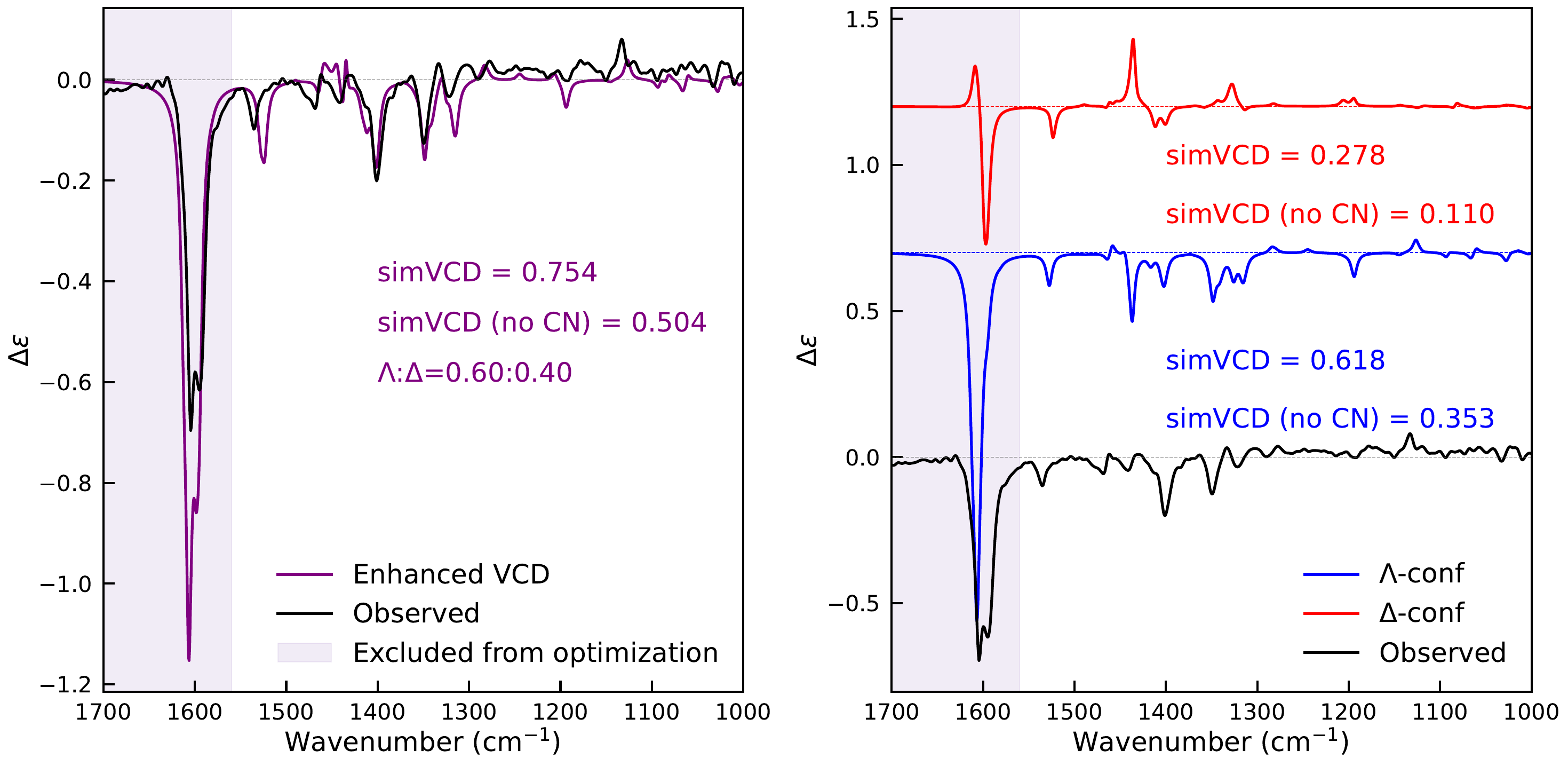}
    \caption{Enhanced VCD spectra averaged over two $\Lambda$ and $\Delta$ conformers (left) for Br as substituent.  Individual weighted contributions of the two conformer with the SimVCD values (right). The fitted excitation energy is 2403 cm$^{-1}$.}
    \label{fig:h_subs}
\end{figure*}
\FloatBarrier

\section{Symmetry analysis}

\begin{table*}[htbp]
\centering
\caption{Symmetry-allowed MFP rotational-strength components and enhancement transitions for chiral point groups $C_n$, $D_n$, T and O ($n$ = 1-6). It indicates which electronic transition can enhance the vibrational modes of particular symmetry and through which Cartesian component of the dipole. For point groups with only proper rotations electric and magnetic dipole moment components transform in the same manner.}
\label{tab:chiral_pg_selection_rules}
\begin{tabular}{cccl}
\hline
Point group & \makecell{VCD/IR \\ Active Mode }& \makecell{Cartesian \\component} & Enhancement transition \\
\hline
 $C_1$ & A  & $x,y,z$ & $A \rightarrow A$ \\
\hline

{$C_2$} & A & $z$  & $A \rightarrow A$, $B \rightarrow B$ \\
& B & $x,y$                            & $A \rightarrow B$ \\
\hline

{$C_3$} & A  & $z$  & $A \rightarrow A$, $E \rightarrow E$ \\
& E  & $x,y$                                  & $A \rightarrow E$ \\
\hline

{$C_4$} & A  & $z$    & $A \rightarrow A$, $B \rightarrow B$ ,  $E \rightarrow E$\\
& E & $x,y$                                 & $A \rightarrow E$, $B \rightarrow E$ \\
\hline

{$C_5$} & A        & $z$  & $A \rightarrow A$, $E_1 \rightarrow E_1$,$E_2 \rightarrow E_2$ \\
& $E_1$  & $x,y$                            & $A \rightarrow E_1$, $E_1 \rightarrow E_2$,  $E_2 \rightarrow E_2$  \\
\hline

{$C_6$} & A        & $z$  & $A \rightarrow A$, $B \rightarrow B$, $E_1 \rightarrow E_1$, $E_2 \rightarrow E_2$ \\
& $E_1$  & $x,y$                                  & $A \rightarrow E_1$, $B \rightarrow E_2$, $E_1 \rightarrow E_2$ \\
\hline

{$D_2$} & $B_1$ & $z$ & $A \rightarrow B_1,\ B_2 \rightarrow B_3$ \\
& $B_2$  & $y$                         &  $A \rightarrow B_2,\ B_1 \rightarrow B_3$  \\
& $B_3$ & $x$                          &   $A \rightarrow B_3,\ B_1 \rightarrow B_2$ \\
\hline

{$D_3$} & $A_2$ & $z$  & $A_1 \rightarrow A_2, E \rightarrow  E$\\
& E        & $x,y$                                  & $A _1\rightarrow E$, $A_2 \rightarrow E$, $E \rightarrow E$ \\
\hline

{$D_4$} & $A_2$  & $z$    & $A_1 \rightarrow A_2, B_1 \rightarrow B_2,  E \rightarrow E$ \\
& E        & $x,y$                                   & $A_1, A_2, B_1, B_2 \rightarrow E$,\\
\hline

{$D_5$} & $A_2$  & $z$   & $A_1\rightarrow A_2$, $E_1 \rightarrow E_1, E_2 \rightarrow E_2$\\
& $E_1$  & $x,y$                                  & $A_1, A_2, E_2 \rightarrow E_1$ \\
\hline

{$D_6$}& $A_2$ & $z$   & $A_1 \rightarrow A_2$, $E_1 \rightarrow E_1$ \\
& $E_1$  & $x,y$                                  & $A_2 \rightarrow E_1$ \\
\hline

T & T       & $x,y,z$     & $A_1, E_1, E_2, T \rightarrow T$ \\
\hline
 O   & $T_1$  & $x,y,z$ & $A_1, E, T_1, T_2 \rightarrow T_1$; $A_2, E, T_2 \rightarrow T_2$\\
\hline
\end{tabular}
\end{table*}

\input{si_fin.bbl}

\bibliographystyle{rsc}

%% file: schiff_complex.bbl
\providecommand*{\mcitethebibliography}{\thebibliography}
\csname @ifundefined\endcsname{endmcitethebibliography}
{\let\endmcitethebibliography\endthebibliography}{}

%% file: si_fin.bbl
\providecommand*{\mcitethebibliography}{\thebibliography}
\csname @ifundefined\endcsname{endmcitethebibliography}
{\let\endmcitethebibliography\endthebibliography}{}